\documentclass[longauth]{aa}  

\usepackage{graphicx}
\usepackage{txfonts}
\usepackage{dblfloatfix}
\usepackage{balance}
\usepackage{float}

\usepackage{hyperref}
\hypersetup{colorlinks,citecolor=blue}
\newcommand{\pycasso}{{\sc p}y{\sc casso}}              
\newcommand{\starlight}{{\sc starlight}}                

\begin{document} 

\titlerunning{The CAVITY project. The spatially-resolved stellar population properties of galaxies in voids}
\title{The CAVITY project. The spatially resolved stellar population properties of galaxies in voids}

\author{Ana M. Conrado \inst{\ref{inst:IAA}}
\and Rosa~M.~Gonz\'alez Delgado\inst{\ref{inst:IAA}}
\and Rub\'en~Garc\'ia-Benito\inst{\ref{inst:IAA}}
\and Isabel~P\'erez\inst{\ref{inst:UGR}, \ref{inst:CarlosI}}
\and Simon~Verley\inst{\ref{inst:UGR}, \ref{inst:CarlosI}}
\and Tom\'as~Ruiz-Lara\inst{\ref{inst:UGR}, \ref{inst:CarlosI}}
\and Laura~S\'anchez-Menguiano\inst{\ref{inst:UGR}, \ref{inst:CarlosI}}
\and Salvador~Duarte Puertas\inst{\ref{inst:UGR}, \ref{inst:CarlosI}, \ref{inst:Canada}}
\and Andoni~Jim\'enez\inst{\ref{inst:UGR}}
\and Jes\'us~Dom\'inguez-G\'omez\inst{\ref{inst:UGR}}
\and Daniel~Espada\inst{\ref{inst:UGR}, \ref{inst:CarlosI}}
\and Mar\'ia~Argudo-Fern\'andez\inst{\ref{inst:UGR}, \ref{inst:CarlosI}}
\and Manuel~Alc\'azar-Laynez\inst{\ref{inst:UGR}}
\and Guillermo~Bl\'azquez-Calero\inst{\ref{inst:IAA}}
\and Bahar~Bidaran\inst{\ref{inst:UGR}}
\and Almudena~Zurita\inst{\ref{inst:UGR}, \ref{inst:CarlosI}}
\and Reynier~Peletier\inst{\ref{inst:Groninguen}}
\and Gloria~Torres-R\'ios\inst{\ref{inst:UGR}}
\and Estrella~Florido\inst{\ref{inst:UGR}, \ref{inst:CarlosI}}
\and M\'onica~Rodr\'iguez Mart\'inez\inst{\ref{inst:UGR}}
\and Ignacio~del Moral-Castro\inst{\ref{inst:Groninguen}, \ref{inst:LaLaguna}, \ref{inst:Chile}}
\and Rien~van de Weygaert\inst{\ref{inst:Groninguen}}
\and Jes\'us~Falc\'on-Barroso\inst{\ref{inst:IAC}, \ref{inst:LaLaguna}}
\and Alejandra Z.~Lugo-Aranda\inst{\ref{inst:Mexico}}
\and Sebasti\'an F.~S\'anchez \inst{\ref{inst:Mexico}}
\and Thijs~van der Hulst \inst{\ref{inst:Groninguen}}
\and H\'el\`ene M.~Courtois\inst{\ref{inst:Lyon}}
\and Anna~Ferr\'e-Mateu\inst{\ref{inst:IAC}, \ref{inst:LaLaguna}, \ref{inst:Australia}}
\and Patricia~S\'anchez-Bl\'azquez\inst{\ref{inst:Madrid}}
\and Javier~Rom\'an\inst{\ref{inst:IAC}, \ref{inst:LaLaguna}}
\and Jes\'us~Aceituno\inst{\ref{inst:IAA}, \ref{inst:CAHA}}
}

\institute{Instituto de Astrof\'isica de Andaluc\'ia (CSIC), PO Box 3004, 18008 Granada, Spain (\email{aconrado@iaa.es})\label{inst:IAA} 
\and Departamento de F\'isica Te\'orica y del Cosmos, Facultad de Ciencias (Edificio Mecenas), Universidad de Granada, E-18071 Granada, Spain \label{inst:UGR}
\and Instituto Carlos I de F\'isica Te\'orica y Computacional, Universidad de Granada, E-18071 Granada, Spain \label{inst:CarlosI}
\and Kapteyn Astronomical Institute, University of Groningen, PO Box 800, 9700 AV Groningen, The Netherlands \label{inst:Groninguen}
\and D\'epartement de Physique, de G\'enie Physique et d’Optique, Universit\'e Laval, and Centre de Recherche en Astrophysique du Qu\'ebec (CRAQ), Québec, QC, G1V 0A6, Canada \label{inst:Canada}
\and Departamento de Astrofísica, Universidad de La Laguna, Avda. Astrofísico Fco. Sánchez s/n, 38200, La Laguna, Tenerife, Spain \label{inst:LaLaguna}
\and Instituto de Astrofísica, Facultad de Física, Pontificia Universidad Católica de Chile, Campus San Joaquín, Av. Vicuña Mackenna 4860, Macul, 7820436, Santiago, Chile \label{inst:Chile}
\and Instituto de Astrofísica de Canarias, Vía Láctea s/n, 38205 La Laguna, Tenerife, Spain \label{inst:IAC}
\and Instituto de Astronom\'ia, Universidad Nacional Auton\'oma de México, A.P. 70-264, 04510, M\'exico, CDMX \label{inst:Mexico}
\and Universit\'e Claude Bernard Lyon 1, IUF, IP2I Lyon, 69622 Villeurbanne, France \label{inst:Lyon}
\and Centre for Astrophysics and Supercomputing, Swinburne University of Technology, Hawthorn VIC 3122, Australia \label{inst:Australia}
\and Departamento de Física de la Tierra y Astrofísica \& IPARCOS, UCM, 28040, Madrid, Spain \label{inst:Madrid}
\and Centro Astronómico Hispano en Andalucía, Observatorio de Calar Alto, Sierra de los Filabres, 04550 Gérgal, Almería, Spain \label{inst:CAHA}
}

\date{Received 31 January 2024 / Accepted 11 April 2024} 

\abstract{The Universe is shaped as a web-like structure, formed by clusters, filaments, and walls that leave large low number-density volumes in between named voids. Galaxies in voids have been found to be of a later type, bluer, less massive, and to have a slower evolution than galaxies in denser environments (filaments and walls). However, the effect of the void environment on their stellar population properties is still unclear. We aim to address this question using 118 optical integral field unit datacubes from the Calar Alto Void Integral-field Treasury surveY (CAVITY), observed with the PMAS/PPaK spectrograph at the 3.5m telescope at the Calar Alto Observatory (Almería, Spain). We fitted their spectra from $3750\AA$ to $7000\AA$ with the non-parametric full spectral fitting code STARLIGHT to estimate their stellar population properties: stellar mass, stellar mass surface density, age, star formation rate (SFR), and specific star formation rate (sSFR). We analysed the results through the global properties, assessing the behaviour of the whole galaxy, and the spatially resolved information, by obtaining the radial profiles from the 2D maps up to the 2 half-light radius of each stellar population property. The results were examined with respect to their morphological type and stellar mass. Then, we compared them with a control sample of galaxies in filaments and walls, selected from the CALIFA survey and analysed  following the same procedure. To make a fair comparison between the samples, we selected a twin filament galaxy for each void galaxy of the same morphological type and closest stellar mass, to match the void galaxy sample as much as possible in morphology and mass. Key findings from our global and spatially resolved analysis include void galaxies having a slightly higher half-light radius (HLR), lower stellar mass surface density, and younger ages across all morphological types, and slightly elevated SFR and sSFR (only significant enough for Sas). Many of these differences appear in the outer parts of spiral galaxies (HLR > 1), where discs are younger and exhibit a higher sSFR compared to galaxies in filaments and walls, indicative of less evolved discs. This trend is also found for early-type spirals, suggesting a slower transition from star-forming to quiescent states in voids. Our analysis indicates that void galaxies, influenced by their surroundings, undergo a more gradual evolution, especially in their outer regions, with a more pronounced effect for low-mass galaxies. We find that below a certain mass threshold, environmental processes play a more influential role in galactic evolution.}
 
   \maketitle

\section{Introduction} \label{Section: introduction}

The large-scale structure (LSS) of the Universe we observe today has evolved since the beginning of time, along with the matter that forms it. During this process, galaxies have born and grown, which is linked to the physical conditions that surround them. Within this context, galaxies in the Universe are not distributed homogeneously: they tend to be located in dense clusters, which are connected by filaments and walls, leaving vast voids in between. Despite their name, the voids are not empty but have galaxies inhabiting them, called void galaxies.

The effect of a high density environment on galaxy evolution has been studied in detail: galaxies in groups and clusters tend to be redder, elliptical, and are forming fewer stars than galaxies in the field \citep{Dressler1980}. 
On the other hand, galaxies in voids have not been as thoroughly investigated, so their properties are not as well known as those of galaxies in higher density environments, such as filaments or walls, where the density is close to the mean density of the universe.

Cosmic voids represent a medium almost unaffected by the complex mechanisms that take place in high density areas, evolving slower and retaining memory of the initial Universe \citep{van_de_Weygaert2011}. For this reason, they are suited to study galaxy formation and evolution, as well as the influence of the LSS on the process. 

Previous studies have found that galaxies in voids are, on average, fainter and less massive \citep{Moorman2015}, bluer and have later-type morphologies \citep{Rojas2004,Kreckel2011,Hoyle2012}, and that they evolve at a slower pace \citep{Dominguez-Gomez2023b} than galaxies located in environments of a higher density. However, when it comes to comparing their stellar formation properties through the star formation rates (SFRs) or specific star formation rates (sSFRs), defined as the SFR per unit stellar mass, a consensus has not been found. 

\citet{Rojas2005} used Sloan Digital Sky Survey (SDSS) spectra to obtain the H$\alpha$ flux as a measure of recent star formation, and found that it was higher for galaxies in voids, independently of their redshift, brightness, and morphological type. On the same line, \citet{Beygu2016A} found that the sSFR, calculated from H$\alpha$ fluxes and near-UV imaging, was also slightly higher in void galaxies. \citet{Moorman2016} also found a higher sSFR for similar stellar mass galaxies in voids, using the same technique. More recent results in \citet{Florez2021} follow the same trend: they find more recent specific star formation in void galaxies, even at a fixed stellar mass and morphological type.

Nevertheless, many other studies have not identified significant differences between galaxies in voids and galaxies in denser environments, specifically among galaxies of the same morphological type. \citet{Ricciardelli2014} took the SFR and stellar mass from the MPA/JHU (Max Planck Institute for Astrophysics/Johns Hopkins University) catalogue \citep{Kauffmann2003, Brinchmann2004, Tremonti2004}, which were derived fitting SDSS spectra with template models, 
and concluded that the environment makes no distinction in the star formation activity, but that quenching happens faster in denser environments (due to the difference in the percentage of passive and star-forming galaxies). \citet{Patiri2006} also found no difference in the sSFR from SDSS spectra at a fixed colour in voids, where they stated that the galactic evolution may not be affected by the global environment but the local one. \citet{Dominguez-gomez2022} used molecular and atomic data, as well as SFRs calculated from measures of aperture-corrected H$\alpha$ fluxes, to conclude that neither the star formation efficiency (SFR divided by the molecular gas mass) nor the sSFR are different in void galaxies. Another approach through the properties of galaxies in the EAGLE (Evolution and Assembly of GaLaxies and their Environments) simulation was carried out in \citet{Rosas-Guevara2022}. They found no discernible variation in the sSFR as a function of stellar mass for star-forming galaxies in different environments.

All of these works studied global properties of galaxies, the general behaviour over their extension inside a certain aperture. However, it  is well known that stellar populations vary spatially, due to the different formation timescales of the different structures within a galaxy, as well as star migration. Galaxies have been shown to grow from the inside out \citep{Perez2013,Gonzalez-Delgado2014,Garcia-Benito2017,Gonzalez-Delgado2017} and stop their star-formation in the same way. The effect of this on the galaxy stellar populations can be seen through negative age radial gradients (e.g. \citealt{Gonzalez-Delgado2015}, \citealt{Parikh2021}, \citealt{Sanchez2020} for a review).

Galaxy-galaxy interactions or mechanisms such as gas pressure stripping can affect the way in which a galaxy forms stars, heating or removing gas in its disc (e.g. \citealt{Gunn1972}). The outer part of galaxies is more prone to be affected by these effects \citep{Koopmann2004,Cortese2012}. For this reason, to assess the effect of the environment on galaxy evolution in a deeper way, it is not enough to study the general behaviour of the galaxies in their full extension, but to see how their properties change spatially as well.

Integral field spectroscopy (IFS) is a powerful observing tool to extract spatially resolved information from the spectrum of different regions in a galaxy. With it, datacubes in three dimensions are obtained (two spatial and one spectral), with which one can get maps of the physical properties that can be derived from galactic spectra. In the last decade, IFS surveys have been extensively used to derive stellar population properties and kinematics of galaxies. Examples of IFS surveys are ATLAS3D (A Three-dimensional Legacy Survey of Massive Galaxies, \citealt{Cappellari2011}), CALIFA (Calar Alto Legacy Integral Field Area, \citealt{Sanchez2012}), SAMI (Sydney-Australian-Astronomical-Observatory Multi-object Integral-Field Spectrograph, \citealt{Croom2012}), or MaNGA (Mapping Nearby Galaxies at APO, \citealt{Bundy2015}).

CALIFA includes a large, homogeneous, but diverse sample of galaxies covering the full Hubble sequence and a wide range of stellar masses \citep{Walcher2014}. They were observed with the PMAS/PPaK IFS \citep{Roth2005, Verheijen2004} at the 3.5m telescope of the Calar Alto Observatory, which can cover the full extent  of these galaxies in its field of view, allowing their properties to be mapped in their whole area. Most CALIFA galaxies are located in filaments and walls (see Section \ref{Subsection: control sample selection}), as it happens in other IFS surveys. None of them are designed to explore galaxies in cosmic voids, and, for this reason, the CAVITY survey was born.

The Calar Alto Void Integral-field Treasury surveY (CAVITY, \citealt{Perez2024}) is a legacy project of the Calar Alto Observatory. It aims to observe the first statistically complete IFU data set of galaxies in voids, making use of the PMAS/PPaK instrument located at the 3.5m telescope of Calar Alto (the same as in CALIFA). The main goals of this project consist in studying the mass assembly (baryonic and dark) and the gas and stellar properties of galaxies in voids to discover the effect of the low density environment on the formation and evolution of galaxies. 

Some studies have already been done based on the CAVITY parent sample (see Section \ref{Section: selection voids} for more details on the sample). Using optical spectra of the CAVITY parent sample from the SDSS, \citet{Dominguez-Gomez2023a} applied a non-parametric SED fitting algorithm to obtain the stellar metallicity of galaxies in three environments: clusters, filaments and walls, and voids. They concluded that galaxies in voids are characterised by lower stellar metallicities with respect to galaxies in filaments and walls, especially for those with a low stellar mass (0.1 dex of difference in the range $10^{9.0}$ to $10^{9.5}$ M$_\odot$), and also with respect to clusters (0.4 dex difference). The latest study is the one in \citet{Dominguez-Gomez2023b}, where, using the same method as in their previous work, they analysed the star formation histories (SFHs) of galaxies in different environments. They found that, on average, the mass assembly of galaxies in voids is slower at a given stellar mass. 

In this work, we enhance our knowledge about void galaxies through the analysis of the datacubes observed by the CAVITY project. We aim to obtain the stellar population properties (stellar masses, ages, SFRs, and sSFRs), for the whole galaxy (with the global quantities) and as a function of the galactocentric radius (through radial profiles). We compare our results to those obtained using the same methodology with datacubes of galaxies in filaments and walls from the CALIFA survey. Our final goal is to find out if the low density environment plays a relevant role in the evolution of galaxies. The stellar population properties (global and in 2D maps) will be available for public access with the CAVITY first data release\footnote{\url{https://cavity.caha.es/}}. The global properties of the galaxies analysed in this work are provided in the Appendix for Table \ref{Tab: tabla anexo}.

This paper is organised as follows. 
For the sake of clarity, we summarise in Sections \ref{Section: observations} and \ref{Section: selection voids} the observations, data reduction, and sample selection, which are explained in detail in the CAVITY presentation paper \citep{Perez2024}. The methodology and the calculations to obtain the stellar population properties are then presented in Section \ref{Section: stellar population analysis}. In Section \ref{Section: results} we present our main results, and we compare them with a control sample of galaxies in filaments and walls in Section \ref{Section: comp CALIFA}. These results are discussed in Section \ref{Section: Discussion}. Finally, Section \ref{Section: Conclusions} summarises the work and the conclusions. The masses calculated for this work were normalised with a Salpeter initial mass function (IMF). Throughout this work we assume a flat $\Lambda$CDM (Lambda cold dark matter) cosmology with $\Omega_M=0.3$, $\Omega_\Lambda=0.7$, and $H_0=70$ km/s/Mpc.

\section{Observations and data reduction} \label{Section: observations}

The observations were obtained with the 3.5m telescope at the Calar Alto observatory, using the Potsdam Multi-Aperture Spectrometer PMAS \citep{Roth2005} in the PPaK mode \citep{Verheijen2004}. PPaK 382 fibre bundle covers a field of view of $74"\times64"$, and each fibre has a diameter of $2.7"$ in the sky. They are distributed in an hexagonal configuration, with a filling factor of ~60\%, with six additional bundles of six fibres each to sample the sky background at the edges of the field of view. The filling factor is raised to 100\% with a three-position dithering pattern. This setup is then translated into datacubes of $78\times73$ pixels, which have an area of $1"\times1"$ each. The datacubes are observed in the low resolution mode V500 ($R\sim850$ at $5000\,\AA$), with a FMHW $\sim6\,\AA$ and typical exposure times of 1.5 or 3 hours, in accordance to the brightness of the galaxies. The covered wavelength range in this mode is $3745-7500\,\AA$ in steps of $2\,\AA$. More details about the sample selection and observation strategy will be presented in \citet{Perez2024}, and related to the data calibration and reduction in Garc\'ia-Benito et al. (in preparation).

The CAVITY pipeline adopts the fundamental reduction methodologies and protocols from the CALIFA survey \citep{Sanchez2016}, while incorporating a new architecture  tailored to the specific requirements of the CAVITY survey.
The reduction process includes the propagation of Poisson and read-out noise as well as the handling of cosmic rays, bad CCD columns, and vignetting.
This process involves combining four FITS files into a single frame, pixel cleaning, fibre tracing, and stray-light correction. After extracting and calibrating spectra, it homogenises them to a common resolution, corrects transmission variations, applies flux calibration, and compensates for atmospheric extinction. Sky subtraction and datacube creation are performed, followed by correction for differential atmospheric refraction and galactic extinction. A more detailed description on the observational and reduction strategy will be provided in the CAVITY presentation paper \citep{Perez2024}.

\section{Sample selection} \label{Section: selection voids}

The CAVITY parent sample consists of 4866 galaxies distributed in 15 voids, carefully selected from the \citet{Pan2012} void galaxy catalogue, which comprises 79947 galaxies from 1055 voids. The 15 CAVITY voids were selected to have different sizes and dynamical stages. This parent sample contains galaxies in the nearby Universe (with redshift values ranging from 0.01 to 0.05). It spans a wide range in stellar masses (from $10^{8.5}$ to $10^{11}$ M$_\odot$), obtained using the mass-to-light ratio from colours technique \citep{McGaugh2014}. These galaxies satisfy the following criteria: they are well located within their respective voids (within 80\% of the void's effective radius), there are at least 20 galaxies distributed within each void, and the distribution of these galaxies in right ascension allows for continuous observability from the Calar Alto observatory throughout the year.

The CAVITY project aims to get IFU observations for a total of $\sim$ 300 galaxies from its parent sample during its 110 awarded useful observing nights. These galaxies do not have too bright stars within the PMAS field, have the suitable size and are not too low surface brightness for this instrument, and have intermediate inclinations. From the ones that were observed until April 2023, we selected those whose quality in the data was enough to perform the spatially resolved analysis (Garc\'ia-Benito, in preparation). We ended up with 109 datacubes of void galaxies from the CAVITY survey. 

In our analysis, we have added nine galaxies from the Void Galaxy Survey (VGS) by \citet{Kreckel2011, Kreckel2012}. These void galaxies were observed using the same instrument and setup as CAVITY during 2019 and 2020, serving as a pilot study. It is worth noting that five of these galaxies are not associated with any of the 15 CAVITY voids, but to others belonging to the void parent catalogue in \citet{Pan2012}, yet we included them in our sample to enlarge it and have better statistics. Consequently, our ultimate selection comprises a total of 118 void galaxies.

For the morphological classification of the CAVITY galaxies, we carried out a cross-match with the morphology catalogue from \citet{Dominguez-Sanchez2018}. To determine the Hubble class, we used the T-Type parameter, which assigns a number to each type of galaxy in a scale from $-3$ (most elliptical) to 10 (irregulars) \citep{deVaucoleurs1963}. Additionally, we take into account the probability of being S0 versus E, p(S0), for the early types (T-Type $<0$), calculated in the same work as the T-Type. The reason we also use this parameter is because of the inefficiency of the Hubble type characterisation model to separate between ellipticals and lenticulars. The p(S0) parameter is obtained using a different model specifically focused on addressing this issue, independent from the one trained to obtain the T-Type parameter, and we use it to make sure we are correctly categorising our galaxies. The correspondence between the parameters from \citet{Dominguez-Sanchez2018} and the Hubble type is described in Table \ref{table:TType-HType}.

\begin{table}[]
\centering
\begin{tabular}{c|cccccc}
       & E              & S0           & Sa  & Sb  & Sc  & Sd  \\ \hline
T-Type & $<0$   & $<0$ & 0 -- 2 & 2 -- 4 &  4 -- 6 &  6 -- 8 \\
p(S0)  & $<0.5$ & $\geq0.5$    &     &     &     &    
\end{tabular}
\caption{Hubble morphological classes as defined from the T type and probability of being S0, p(S0).}
\label{table:TType-HType}
\end{table}

\begin{figure}
\centering
\includegraphics[width=0.45\textwidth]{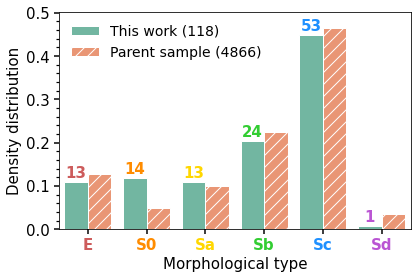}
\caption{Density distribution of morphological types in the CAVITY parent sample (4866 galaxies) and the selected ones (118 galaxies). The numbers above the bars indicate the number of galaxies in the selected sample of each morphological type.}
\label{fig:htype}
\end{figure}

Figure \ref{fig:htype} shows the distribution of morphological types for the parent sample and the chosen subsample. The coloured numbers on top of the green bars determine the number of galaxies of each type that we have in the subsample. As expected from what was seen in previous works \citep{Rojas2004,Kreckel2011,Hoyle2012,Beygu2016A}, most of the galaxies in voids belong to later types (Sbs and Scs). Our selection follows the same trend as the parent sample and has at least 13 galaxies of each type, with the exception of Sds, where we found only one. For statistical significance reasons, we exclude the Sd from those results where we stack the calculated properties by morphological type (in Sections \ref{Section: results} and \ref{Section: comp CALIFA}).

\begin{figure}
\centering
\includegraphics[width=0.45\textwidth]{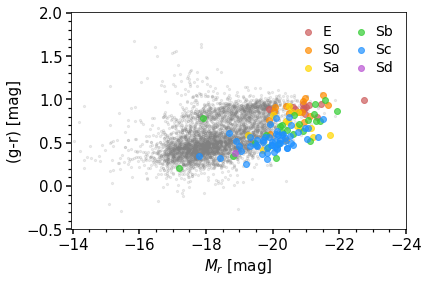}
\caption{Comparison between the full CAVITY parent sample for galaxies with available Hubble types in \citet{Dominguez-Sanchez2018} (grey points) and the selected galaxies (coloured by morphological type) in the colour-magnitude diagram.} 
\label{fig:col-mag}
\end{figure}

The colour-magnitude diagram of the parent sample and the selection for this work is shown in Figure \ref{fig:col-mag}. Although the parent sample covers a wider range in absolute magnitude in the r band ($M_r$,) our subsample is more restricted to the brightest part of the diagram. This is due to the selection criteria: we choose the brightest galaxies in order to have enough signal-to-noise ratio (S/N) to perform the fit reliably out to large galactocentric distances and obtain the spatially resolved information. The observational strategy also takes part in this matter: the brightest galaxies are the ones that have been observed first. We must keep this in mind when discussing the results, as this bias could compromise the generalisation of our results for galaxies in voids.

\section{Stellar population properties analysis} \label{Section: stellar population analysis}

\subsection{Method of analysis} \label{Subsection:method}

We make use of full spectral fitting techniques in order to obtain information about the stellar populations that contribute to a given galaxy spectrum. We use \starlight \ \citep{CidFernandes2005}, a non-parametric spectral synthesis code that fits all the stellar features of the galaxy spectrum
with a superposition of spectra of single stellar populations (SSPs). These SSPs are the spectra of groups of stars of a certain age and metallicity. They can be obtained from stellar libraries, empirical or synthetic. The collection of SSPs that we use in our analysis is a combination of both types. First, for ages $\geq$ 63 Myr, we consider \citet{Vazdekis2010, Vazdekis2015} SSPs, based on the empirical stellar library MILES, with 23 ages logarithmically distributed between 0.06 Myr and 14 Gyr and eight metallicity values ($Z=0.0001,0.0003,0.0010,0.0040,0.0080,0.0220,0.0340,0.0470$). These SSPs have a spectral range of 3540–7409 \AA, with a resolution of $2.3\,\AA\ $. We complement them with the synthetic models from \citet{Gonzalez-Delgado2005} for the younger populations (< 63 Myr). These models cover a wavelength range between 3000 and 7000 \AA, and have a resolution that is degraded to be equal to that of MILES. We choose a set of 5 ages (between 1 and 5.6 Myr) and 5 metallicites (between 0.0001 and 0.0470), both distributed uniformly in logarithmic scale. Both collections of SSPs use Salpeter as their IMF, a fixed [$\alpha$/Fe = 0], and the isochrones used are those of \citet{Girardi2000} except for ages 1 and 3 Myr, that use the Geneva tracks \citep{Schaller1992,Schaerer1993,Charbonnel1993}.

Given that \starlight \ only fits the stellar continuum and absorption lines of the spectrum, we mask the nebular emission lines. Additionally, we must adjust the spectral resolution of the SSPs to match that of the observed data, achieved by convolving them with a Gaussian kernel. This method was used and explained in previous works \citep{Perez2013,Gonzalez-Delgado2014,Gonzalez-Delgado2015}. 
 Once we have prepared our bases and applied the necessary masks and resolution adjustments, we can proceed to fit the spectra. We discuss the quality of the fits in Appendix \ref{section: appendix 2D maps}.

\subsection{Stellar population properties} \label{subsection calculations}

With \starlight \ we can obtain information of the SFHs of galaxies, parametrised by the light population vector ($x_{t,Z}$). This vector contains the percentage that each SSP is contributing to the total continuum light of the spectrum. Through the mass-to-light ratio of each SSP, we can estimate the mass population vector ($\mu_{t,Z}$), which is similar to $x_{t,Z}$ but referring to the stellar mass fraction due to each element.

The present stellar mass is calculated from the SFH as the mass converted into stars, taking into account the mass loss of the SSP due to stellar evolution. We use these measurements of the stellar mass to perform our calculations.

With the population vectors, we can calculate the light or mass-weighted logarithmic age, representative for the entire spectrum, summing the age of each SSP multiplied by their corresponding value of the $x$ or $\mu$ vectors, after normalisation. We choose to compute the weighted median of the logarithm of the age instead of just taking it in linear scale for robustness reasons \citep{Gonzalez-Delgado2015}. They are calculated as:

\begin{equation}
    \langle \text{log t} \rangle _L=\sum_{t,Z}x_{t,Z} \ \text{log t};
\end{equation}

\begin{equation}
    \langle \text{log t} \rangle _M=\sum_{t,Z}\mu_{t,Z} \ \text{log t}.
\end{equation}

In this study, we assess the age of the galaxies through the light-weighted ages. The reason for this choice is because the contribution from younger populations, which is important when calculating the mean age, is mapped better by their light than by their mass \citep{Gonzalez-Delgado2015, Garcia-Benito2017}.

We determine the current SFR by summing the stellar mass formed over the latest time interval and dividing it by its duration (see Equation \ref{eq:SFR}). We define a star-forming age, $t_{\text{SF}}$ (the age of the older stars taken into account in the sum), as an arbitrary value for this calculation. 

Selecting a higher value for $t_{\text{SF}}$ would yield a more robust result, as it includes a greater number of terms in the calculation. However, this may diminish the sensitivity to recent changes. Conversely, opting for a smaller $t_{\text{SF}}$ would result in less reliable outcomes, making the results more susceptible to errors and degeneracies. In this work we take $t_{\text{SF}}=32$ Myr, as it was found to give results that correlate with EW(H$\alpha$), the equivalent width in H$\alpha$ \citep{Gonzalez-Delgado2016}. We calculate the SFR by:

\begin{equation} \label{eq:SFR}
    \text{SFR}(t_{\text{SF}})=\frac{1}{t_{\text{SF}}}\sum_{t \leq t_{\text{SF}}}\text{M}(t).
\end{equation}

Occasionally, the process may yield a minor presence of a young population component in early-type galaxies. This outcome may not be a genuine representation but rather a residual value resulting from the optimisation algorithm. \citet{CidFernandes2014} estimated the amplitude of the uncertainties that \starlight \ can introduce, and concluded that the level of noise in the young population contribution in light $x_Y$ (the sum of those younger than $t_{\text{SF}}$) is around 3\%. These values are low, but when multiplying by the total mass of the galaxy to denormalise, the value can get high enough for massive galaxies to make a difference in the calculation of the recent SFR. For this reason, we mask the spaxels whose $x_Y$ do not reach a certain limit. The choice of this value was taken from \citet{Gonzalez-Delgado2016}, that calculated the correlation between $x_Y$ and EW(H$\alpha)$, based on CALIFA data. \citet{Stasinska2008} and \citet{CidFernandes2011} found that the limit of EW(H$\alpha)$ that a 'retired' galaxy (one that is not forming stars) can have due to hot, old, low mass stars is EW(H$\alpha)=3\,\AA$, which corresponds to a population fraction $x_Y=3.4\%$, so we consider this limit value.

The sSFR is the SFR per unit stellar mass, so we obtain it by dividing the SFR by the total present stellar mass in the area of the galaxy that encompasses the spectrum:

\begin{equation} \label{eq:sSFR}
    \text{sSFR}(t_{\text{SF}})=\frac{\text{SFR}(t_{\text{SF}})}{\text{M}_\star}.
\end{equation}

Because we focus in the radial structure of the SFR and sSFR, we calculate the intensity of the SFR, also called the SFR surface density, which is the SFR per unit area, $\Sigma_{\text{SFR}}$. With it, we can calculate the local specific SFR as sSFR = $\Sigma_{\text{SFR}}/\Sigma_\star$, being $\Sigma_\star$ the stellar mass surface density (the mass per unit area).

There are other magnitudes that are modelled during the fit, as the stellar extinction (A\textsubscript{V}). We choose the dust reddening law from \citet{Cardelli1989} with R\textsubscript{V} = 3.1.

\subsection{Half-light radius} \label{subsection: HLR}

To ensure that our comparison of spatially resolved properties remains independent of a galaxy's physical size or its apparent size in the sky, we employ the half-light radius (HLR) as a normalisation measure. The HLR is a metric that takes into account the radius at which half of the total luminosity of a galaxy is contained. It is calculated using elliptical annuli, considering both the position angle and ellipticity, which are derived from the flux map in the photometric r-band from the original datacube. The centre is previously calculated as the brightest pixel in the central part for the galaxy. We use the unbinned datacube rather than the segmented one (i.e. Voronoi zones), as explained in subsection \ref{subsection:spat properties}. This approach also allows us to determine the position angle and ellipticity of each galaxy.

Another parameter that can also be used instead of the HLR is the half-mass-radius (HMR). It is equivalent to the HLR, but is calculated as the radius that contains half of the total stellar mass, instead of the luminosity. Previous studies have shown that the ratio between the HLR and the HMR is not constant, but changes with the mass and morphology of galaxies \citep{Gonzalez-Delgado2015}. Depending on the nature of the study one can choose to use one or the other, understanding that they are not equivalent. For this work, we decide to use the HLR, as it is a more straightforward measurement that can be obtained from the original datacubes.

Our HLRs span a range of 1.5 to 10.6 kpc, with a mean value of 4.1 kpc and dispersion 1.4 kpc. We find that later types (Sbs and Scs) have larger HLR than early types. This can be due to a larger outer radius, or to a lower concentration of light in the centre, due to differences in their intrinsic light distribution. 

\subsection{Spatially resolved properties} \label{subsection:spat properties}

Our data consist of IFU datacubes, so we need a way to fit the spectrum of every individual spaxel and then analyse everything together. To do so, we use \pycasso : a pipeline that preprocesses the data, creates an output datacube and provides tools for its analysis \citep{de_Amorim2017}.

The preprocessing stage consists of various steps. First, we mask those spaxels with a S/N lower than a threshold value (we choose S/N$_{\text{threshold}}=3$ per spaxel). The S/N is calculated in the rest-framed spectral range $5650 \pm 60 \,\AA$, as the signal, the median value inside the wavelength window, divided by the noise, the standard deviation of the detrended flux in the same range. We can also apply a spatial mask in case we have foreground stars or spurious sources. Then, to make sure we have enough S/N in every spectrum to perform a reliable full spectral fitting, the datacubes are segmented into Voronoi zones \citep{Cappellari2003}. We select S/N$_{\text{target}}=20$ per zone. The spectra is then rest-framed and resampled to the range $3750-7000$ \AA, with $\Delta \lambda$ = 2 \AA. Some plots including quality control maps that are obtained after the fit are shown in Appendix \ref{section: appendix 2D maps}.

After the preprocessing stage, \pycasso \ takes every individual Voronoi zone's spectrum and feeds it to \starlight \ for the full spectral fitting. Then, the individual output files are stored in a multi-extension binary FITS file to easily access the maps of each \starlight \ output. With these we can already evaluate the spatially resolved properties of a galaxy, and obtain the 2D maps of the desired properties. The \starlight \ output of extensive properties, such as the luminosity or the mass, depends on the normalisation of the input spectrum. For this reason, we must divide the output of a certain zone by the number of spaxels that have been summed to obtain the value in each pixel of the zone. On the contrary, intensive properties, as the age, are not related to the normalisation of the input spectrum, and thus the result obtained by the fit is the one that will be shared in every pixel of each zone.

\begin{figure*}
\centering
\includegraphics[width=\textwidth]{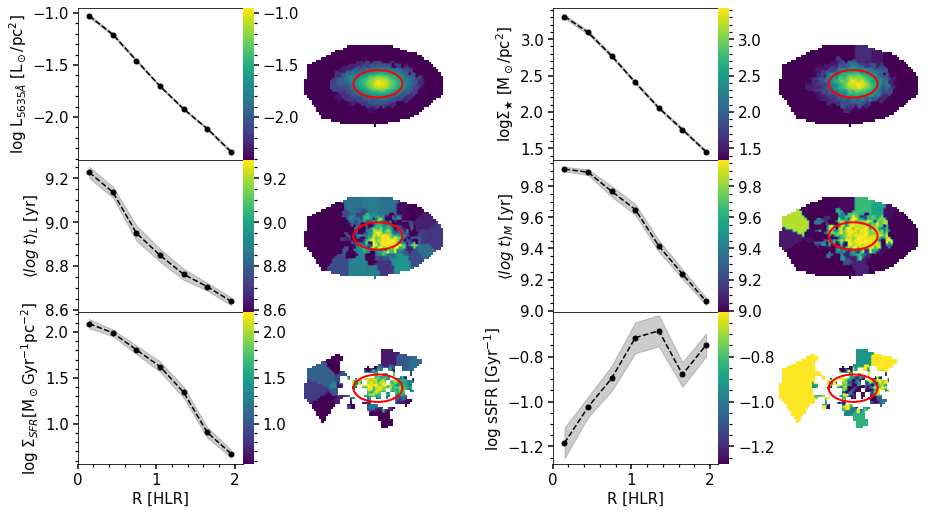}
\caption{Radial profiles and maps of some physical properties of CAVITY54706, one of the Sc galaxies in our sample. The red ellipse in the maps indicates its half-light radius. The colour bars share scale with their corresponding profile's y-axis. In the left and middle left column, from top to bottom: Luminosity at the normalisation wavelength ($\lambda=5635\AA$), light-weighted ages, and intensity of the SFR (SFR surface density). In the right and middle right column, from top to bottom: stellar mass surface density, mass-weighted ages, and specific SFR.}
\label{fig:maps_profiles}
\end{figure*}

From the 2D maps, we can also get the radial profiles (assuming axisymmetry). Starting at the centre of the galaxy, we take elliptical apertures with the same position angle and ellipticity as the galaxy in growing intervals of radius over HLR (R/HLR). The median value of the pixels that are in each section gives the value of the profile at each distance to the centre, and we can estimate the error by doing the same with the standard deviation. Figure \ref{fig:maps_profiles} shows the maps and corresponding profiles of different properties of an example void galaxy. In the maps, it is possible to see the Voronoi zones in the outer parts, where many pixels share the same values.

We stack the radial profiles by morphological type for an easier interpretation (with the exception of the only Sds, due to low statistics), as well as show them in categories of mass and morphology. We make two morphology bins: early (Es and S0s) and spiral (Sbs and Scs) types; and two stellar mass bins: low mass (log M$_\star/$M$_\odot\leq10.5$) and high mass (log M$_\star/$M$_\odot>10.5$). The reason why we do not add the Sas in any of the two categories (early or spiral) is because they usually show values in between of early-type and spirals, and we wanted to separate the two groups as cleanly as possible. The limit value for the stellar mass categories is the median value of the total stellar mass of the selected sample. Table \ref{table:4-cat profiles} shows the number of galaxies that belong to each of the four categories. The low mass, early type category only has four galaxies, so we must keep it in mind when looking at the results, as they could be unreliable.

\begin{table}[]
\centering
\begin{tabular}{c|cc}
              & Low mass & High mass \\ \hline
Early (E, S0) & 4        & 23        \\
Spiral (Sb, Sc) & 59       & 18       
\end{tabular}
\caption{Number of galaxies in each category for the mass-morphology, segregated and stacked radial profiles. The separation between low and high mass is done at the median value for all galaxies, log M$_{\star, \text{median}}=10.5$ [M$_\odot$]. The total stellar mass ranges between $1.0\times10^9$ and $4.2\times10^{11}\,$M$_\odot$.}
\label{table:4-cat profiles}
\end{table}

\subsection{Global properties}

To assess the overall behaviour of a galaxy, we obtain its global properties. Extensive quantities such as the stellar mass can be directly calculated by summing over the spatial zones. However, intensive properties like the age, must be derived by fitting the integrated spectrum. This integrated spectrum is generated by summing the spectra of every unmasked spaxel. The area covered by the integrated spectrum is equal to the sum of the areas of the Voronoi zones, and we take it as the total area of the galaxy. We analyse the global properties looking at their distribution by morphological type or in a scatter plot against the total stellar mass.

\section{Stellar populations properties of galaxies in CAVITY} \label{Section: results}

In this section, we present the global and spatially resolved results for each stellar population property analysed (stellar mass surface density, age, SFR and sSFR). The values of the global median values and standard deviations for each morphological type are presented in Table \ref{tab:median_global_results}, at the end of the section.

\subsection{Stellar mass}

\begin{figure}
\centering
\includegraphics[width=0.45\textwidth]{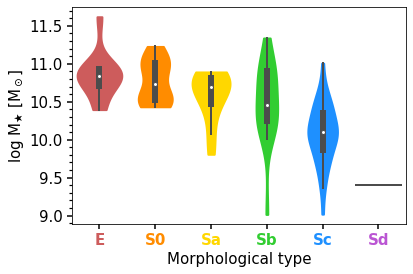}
\caption{Violin plots of the logarithm of the stellar mass for each morphological type. The coloured area determines the density distribution (Gaussian kernel), the white dot inside each of them marks the median value of each category, the thicker dark line ends in the first and third quartiles, and the thinner line reaches until the lower and upper adjacent values (points outside its range are considered outliers). The edges of the distributions are cut to define the minimum and maximum values. The Sd category only shows a line with the value for its single galaxy.}
\label{fig:mass_morph}
\end{figure}

The distribution of stellar masses (M$_\star$), obtained from the spectral fitting, for each morphological type is shown in Figure \ref{fig:mass_morph}. The mass ranges between $1.0\times10^9$ and $4.2\times10^{11}\,$M$_\odot$, having its median at $2.6\times10^{10}\,$M$_\odot$, and it can be seen that it is slightly correlated with the Hubble type as expected, given that we are not including dwarf galaxies: earlier types tend to be more massive than later types. Sbs have the largest dispersion, with values ranging from $10^9$ to $10^{11.3}\,$M$_\odot$.

\subsection{Stellar mass surface density}

\begin{figure}
\centering
\includegraphics[width=0.45\textwidth]{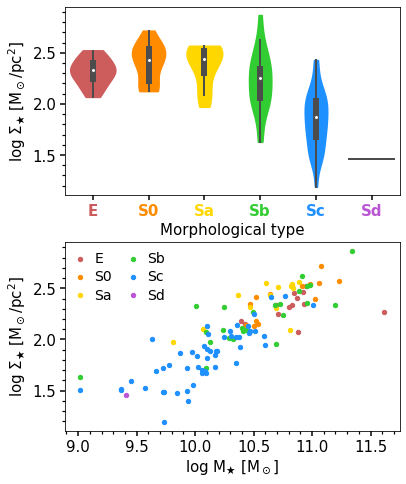}
\caption{Global stellar mass density coloured by morphological type.  \textit{Top}: Distribution of the $\Sigma_\star$ for each type (violin plot as in Figure \ref{fig:mass_morph}). \textit{Bottom}: Stellar mass density versus stellar mass.}
\label{fig:integ_Mdens}
\end{figure}

From the total stellar mass, we calculate the stellar mass surface density $\Sigma_\star$ = M$_\star$/area by dividing the stellar mass by the total area of the galaxy (for the global values) or the area in each zone (for the maps and the radial profiles). Figure \ref{fig:integ_Mdens} shows the global $ \Sigma_\star $ for each morphological type, its distribution on the top panel and the scatter plot versus the total stellar mass on the bottom. Comparing the top panel with the distribution of the total stellar mass (in Figure \ref{fig:mass_morph}), one can see that most morphological types follow the same distribution and relative values except for ellipticals, which have lower $\Sigma_\star$ than lenticulars and even Sas. The shape of the distribution in the S0s varies, having more weight towards the denser direction, but the rest of the types have similar distributions for M$_\star$ and $\Sigma_\star$.
 
In the bottom panel, the scatter points form a sequence of growing $\Sigma_\star$ with M$_\star$. As expected, later types (Sc and Sd), that are usually less massive, populate the lower part of this relation, while earlier types are in the higher end. We do not find an abrupt change in the slope of the relation at the boundary between spiral and early types, as in \citet{Kauffmann2003} for SDSS galaxies. Instead, our results are more similar to those in \citet{Gonzalez-Delgado2015}, where they found a smoother transition using the CALIFA sample.

\begin{figure}
\centering
\includegraphics[width=0.45\textwidth]{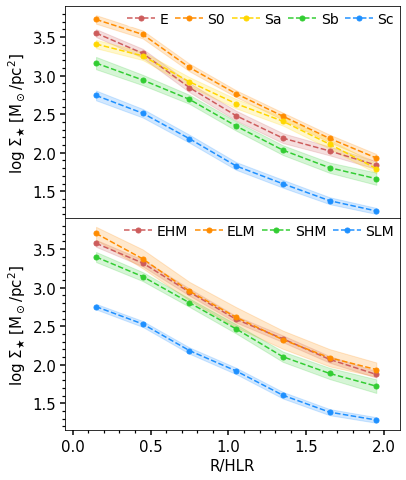}
\caption{Stacked radial profiles of the stellar mass density. The shaded area marks the uncertainty, calculated as the standard deviation divided by the square root of the number of points in each R/HLR bin. \textit{Top}: Profiles by morphological types. We chose not to plot the profile from the Sd galaxy, as there is only one.  \textit{Bottom}: Profiles by morphological and mass bins. We made 4 categories: early (E and S0) and spiral (Sb and Sc), and low (log M$_\star/$M$_\odot\leq10.5$) and high (log M$_\star/$M$_\odot>10.5$) mass. The number of galaxies in each category is described in Table \ref{table:4-cat profiles}. The acronyms stand for early type high mass (EHM), early type low mass (ELM), spiral high mass (SHM) and spiral low mass (SLM).}
\label{fig:prof_Mdens}
\end{figure}

The stacked radial profiles for galaxies of the same morphological type are shown in Figure \ref{fig:prof_Mdens}. The coloured area around each line marks the error
of the mean, calculated as the standard deviation divided by the square root of the amount of points at each radius from the centre. The top panel displays the profiles of each morphological type (excluding the Sd). All the morphological types show decreasing $\Sigma_\star$ with galactocentric radius. We find that the profiles scale with morphology, having higher values for earlier types and lower for later types, with the largest separation between Sc and Sb. An exception to this is found in the ellipticals, whose profile reach lower values than S0s at all galactocentric distances and Sas everywhere but in the innermost region. This is in disagreement with previous studies done for the CALIFA sample \citep{Gonzalez-Delgado2015}. We suspect the reason of this behaviour could be due to the selection of the galaxies, and that this may change for a larger sample.

The bottom panel of Figure \ref{fig:prof_Mdens} shows the profiles stacked in categories of morphological type (early and spiral) and mass. We exclude Sas from the two categories, as we wanted a clear distinction between the two types of galaxies (early and spiral). The four groups are early high mass (EHM), early low mass (ELM), spiral high mass (SHM) and spiral low mass (SLM). A previous study \citep{Gonzalez-Delgado2015} found a clear scaling of the radial profiles of $\Sigma_\star$ with the stellar mass, but we do not find such an evident behaviour: a remarkable difference can be seen between spiral type, low mass galaxies, which stay way lower, but the rest, independently of their mass or type, stay very close to each other. We find much more similar values when we separate early types by mass (in the bottom panel), which values of log $\Sigma_\star$ [M$_\odot$/pc$^2$] at R = 1 HLR are $2.62\pm0.13$ and $2.59\pm0.04$ for low and high mass, respectively, than when we do it by morphological type (top panel), that have values of $2.48\pm0.06$ for E and $2.77\pm0.04$ for S0 at the same galactocentric distance (all in M$_\odot$/pc$^2$ in logarithmic scale).

\subsection{Stellar ages}

\begin{figure}
\centering
\includegraphics[width=0.45\textwidth]{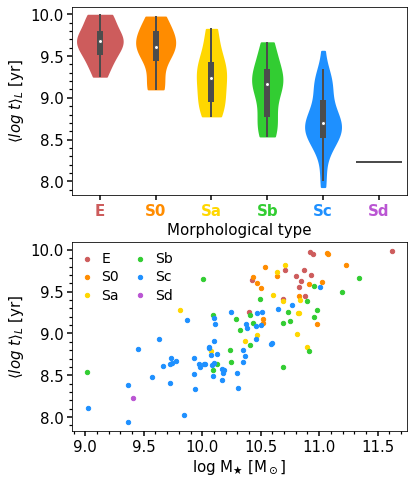}
\caption{Global light-weighted ages. \textit{Top}: Distribution of $\langle \text{log t} \rangle _L$ for each type (violin plot as in Figure \ref{fig:mass_morph}). \textit{Bottom}: Light-weighted ages versus stellar mass.}
\label{fig:integ_logt}
\end{figure}

The top panel in Figure \ref{fig:integ_logt} shows the distribution of the stellar age derived from the integrated spectrum for each morphological type. There is a clear increase of $\langle \text{log t}\rangle _L$ from the Sd to the Sa galaxies. Earlier types, E and S0, have similar ages, older than Sas. This trend is similar to that found for the CALIFA sample in \citet{Gonzalez-Delgado2015}. The bottom panel shows an increasing sequence of the average age with the galaxy stellar mass, reflecting the downsizing behaviour of the stellar populations that was shown in previous works \citep{Gallazzi2005, Neistein2006, Gallazzi2021}. The dispersion of this M$_\star$-age relation is also related to galaxy morphology. One can see that although the Sbs reach high values of M$_\star$, for a fixed mass, earlier types are older.

\begin{figure}
\centering
\includegraphics[width=0.45\textwidth]{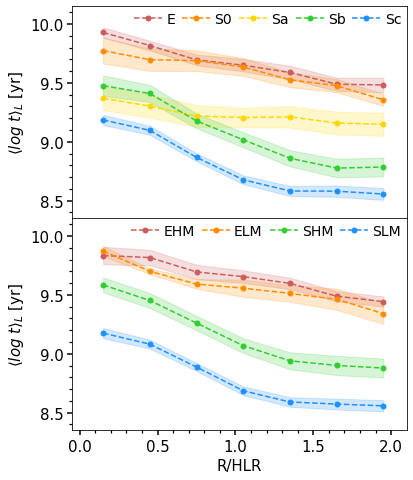}
\caption{Radial profiles of the light-weighted ages. \textit{Top}: Stacked profiles by morphological type.  \textit{Bottom}: Stacked profiles by morphological and mass bins. See Figure \ref{fig:prof_Mdens} for more details.}
\label{fig:prof_logt}
\end{figure}

Figure \ref{fig:prof_logt} shows the age radial profiles, stacked by morphology (upper panel) and by stellar mass and morphology (bottom panel). They show the expected decrease of the age from the centre towards larger distances, reflecting the inside-out formation scenario of galaxies (e.g. \citealt{Perez2013, Sanchez-Blazquez2014, Gonzalez-Delgado2014, Gonzalez-Delgado2017, Garcia-Benito2017, Goddard2017}). It is worth noting that this scenario is not universal for all kinds of galaxies, but is restricted to certain types and stellar mass ranges. Very massive early type galaxies ($\text{M}\sim10^{12}\,\text{M}_\odot$), usually those in the centre of clusters, are found to have flat age profiles \citep{Oliva-Altamirano2015}, while dwarf ellipticals ($\text{M}<10^{9}\,\text{M}_\odot$) tend to have positive age gradients \citep{Bidaran2023}. However, none of the galaxies analysed in this work lay on any of these groups.

The radial profiles also scale with morphology, being the S0 and E older at every galactocentric radius than the spirals, and Scs being younger than Sas and Sbs everywhere as well. However, we find that the inner parts of Sas have an age similar to the Sbs. \citet{Falcon-Barroso2006} saw that Sa galaxies were more star-bursting in their bulges, probably because of the presence of star-forming rings. This behaviour can explain the younger ages of the Sa central parts, reaching values similar to or even lower than Sbs.

The radial profiles of the light-weighted age also show a clear dependence with M$_\star$, as seen in the bottom panel of Figure \ref{fig:prof_logt}. The most massive spirals are older than the low-mass spirals at every radius. The profiles of the early types are older than those of the spirals at every galactocentric distance, but there is not that much difference between high and low mass for early types. We find that the slope of the profiles behaves differently between the two morphological bins: it is steeper for spirals, where the younger stars mainly live in their discs, and flatter for early types, which have their stellar populations more homogeneously distributed.

\subsection{SFR and intensity of the SFR} \label{subsection: results SFR}

\begin{figure}
\centering
\includegraphics[width=0.45\textwidth]{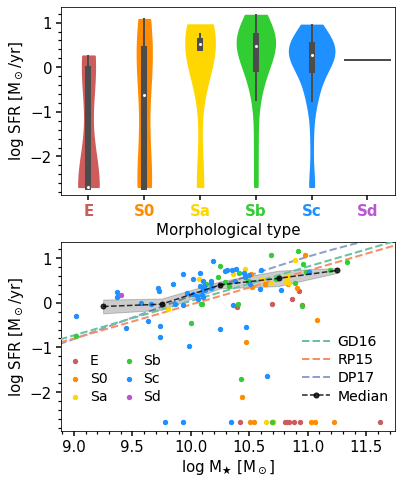}
\caption{Global SFR. The galaxies with SFR = 0 M$_\odot$/yr were given the value SFR = 10$^{-2.67}$ M$_\odot$/yr (the value of the minimum minus the standard deviation) so that they could appear in the plots. \textit{Top}: Distribution of the SFR by morphological types (violin plot as in Figure \ref{fig:mass_morph}). \textit{Bottom}: SFR versus stellar mass, coloured by morphological type. The dashed lines refer to the SFMS, parametrised as in \citet[][in green]{Gonzalez-Delgado2016}, \citet[][in orange]{Renzini_Peng2015} and \citet[][in blue]{Duarte-Puertas2017}. The black dots mark the median value of our data in bins of mass, taking into account only late spirals (Sb, Sc, Sd) that had originally SFR $\neq$ 0.
}
\label{fig:integ_SFR}
\end{figure}

The global SFR for each galaxy is calculated using Equation \ref{eq:SFR}, adding the contribution of all non-masked spaxels. If the light contribution of the spatial sum of the young populations is lower than the limit value $x_Y=3.4\%$, we set the SFR = 0, as explained in Section \ref{subsection calculations}. Figure \ref{fig:integ_SFR} (top panel) shows the distribution of SFR for each morphological type. The galaxies that have not formed stars significantly in the last 32 Myr (SFR = 0) would be masked, but we gave them a representative value of SFR = 10$^{-2.67}$ M$_\odot$/yr, to keep the quenched galaxies in the diagram. This value is the minimum of all the galaxies minus the standard deviation, chosen as the minimum reached value taking into account the uncertainty. If we did not do this, it would seem that the analysed galaxies (especially early types) have way higher SFR on average. We find that most of ellipticals (8 out of 13) and some S0 (5 out of 14) do not have relevant SFR. The fraction of galaxies with SFR = 0 decreases with later morphological types, being $62\%$, $36\%$, $15\%$, $4\%$, and $5\%$ for E, S0, Sa, Sb, and Sc galaxies, respectively.

The bottom panel in Figure \ref{fig:integ_SFR} has the SFR-M$_\star$ relation, also called the star-formation main sequence (SFMS, \citealt{Noeske2007}), which has been found to be a universal scaling law, followed by star-forming galaxies up to $z\sim 4$ \citep{Schreiber2015, Speagle2014}. Galaxies whose star formation is diminishing will leave the sequence through the green valley to the area where quiescent galaxies with very low SFR lie. Different parametrisations of the SFMS are represented in the panel with dashed lines of different colours. These are obtained from \citet{Gonzalez-Delgado2016} (fitted over Sc galaxies from the CALIFA sample), \citet{Renzini_Peng2015} and \citet{Duarte-Puertas2017} (that fitted a 3rd-degree polynomial). Over them, we show the median values of our late spiral galaxies (Sb, Sc, and Sd) in bins of mass. While calculating the median value, we did not take into account those galaxies that originally had SFR = 0. We did not perform a linear fitting due to the low number of galaxies.
We find that star-forming galaxies in voids follow the SFMS. Some of our galaxies deviate from the general rule that says that early-type galaxies are quiescent and out of the SFMS (38\% of our elliptical galaxies are forming stars) and that late types follow the sequence (we have three quenched Sc galaxies). This result was also found in previous works such as \citet{Vulcani2015}, \citet{Gomes2016}, \citet{Bitsakis2019}, \citet{Cano-Diaz2019}, and \citet{Paspaliaris2023}, where they saw a fraction of early type galaxies forming stars and late type galaxies below the SFMS in different environments.

\begin{figure}
\centering
\includegraphics[width=0.45\textwidth]{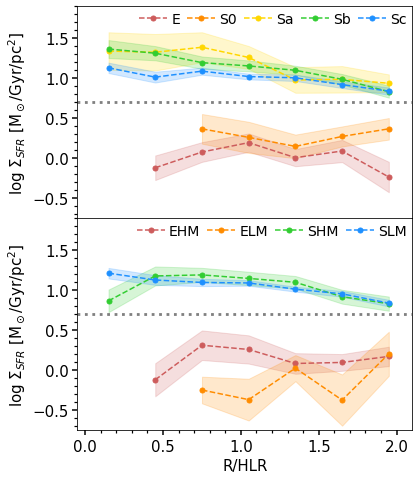}
\caption{Radial profiles of the intensity of the SFR or the SFR surface density. The dotted grey line marks the mean SFR density in the Milky Way \citep{Licquia2015}. \textit{Top}: Stacked radial profiles by morphological type. \textit{Bottom}: Stacked radial profiles in bins of morphology and mass. The reason why the profiles that belong to early types are cut in their inner regions is because of the masking: the spaxels in their centre did not have recent star formation, or it was not significant enough (it did not surpass the limit of the 3.4\% contribution in light). See Figure \ref{fig:prof_Mdens} for more details.}
\label{fig:prof_SFRdens}
\end{figure}

The profiles of the intensity of the SFR (the SFR per area, $\Sigma_{\text{SFR}}$) are shown in Figure \ref{fig:prof_SFRdens}, stacked by morphological types on the top panel and by bins of morphology and mass at the bottom. Galaxies that were found to have a global SFR = 0 may be forming stars in some of their regions (usually in their outer parts), while their quenched areas are masked. If the area of these regions is small compared to the area of the whole galaxy, the weight in light of the young populations once summed up with the rest of the spatial regions can end up being lower than the threshold value (3.4\%), and therefore resulting in a global SFR = 0. For this reason, when computing the median profiles, we take every galaxy into account. We can distinguish two different behaviours: ellipticals and S0s have flat profiles with low values, below the reference line at all galactocentric distances, while spirals overlap in a decreasing trend above the line. This is because the spirals are still forming stars at a rate that follows the SFMS, while earlier types have already left the main sequence and have almost no recent SFR. The reference line marks the mean SFR of the Milky Way, $\Sigma_{\text{SFR,MW}}$ = 5 M$_\odot $Gyr$^{-1}$pc$^{-2}$ \citep{Licquia2015}, calculated as the division between the Milky Way's recent SFR and the area of a circle of radius 1.2 times the galactocentric radius of the Sun, which is $\sim$ 2 HLR. 

The reason why galaxies that are in the main sequence have profiles with similar values is because of the existing scaling relation between the SFR and the stellar mass (the SFMS) and between the size of a galaxy and the stellar mass. The connection between these two relations results in an invariance for the profiles of the intensity of the SFR with stellar mass. The median decreasing of the log $\Sigma_{\text{SFR}}$ per HLR is $-0.066$ [M$_\odot$/yr/pc$^2]$.

The profiles in the bottom panel of Figure \ref{fig:prof_SFRdens} show the same behaviour: flat profiles with lower values for early types that are mainly quiescent, and a negative gradient for spirals. We note that, as expected, the stellar mass does not play any differentiating role. 

\begin{figure*}
\centering
\includegraphics[width=\textwidth]{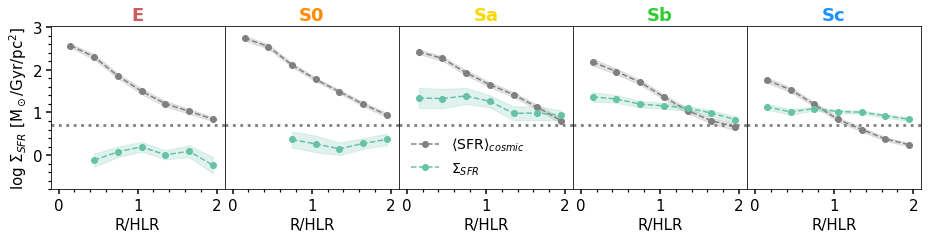}
\caption{Profiles of the intensity of the SFR, stacked by morphological type in each panel. The green lines correspond to the ones obtained observationally, and the grey ones to the mean cosmic SFR: the profiles that we would obtain for the galaxy to form all its stellar mass if it had had a fixed SFR since the beginning of the Universe. The dotted grey line marks the mean SFR density in the Milky Way \citep{Licquia2015}.}
\label{fig:cosmic_SFRdens}
\end{figure*}

The low values of $\Sigma_{\text{SFR}}$ of E and S0 galaxies are clear evidence that the star formation has been quenched. We can compare with the profiles that would be obtained assuming a fixed SFR since the beginning of the Universe, $\langle \Sigma_{\text{SFR}}(R)\rangle _{cosmic}\sim \Sigma_\star(R)/t_{univ}$, taking into account that this $\Sigma_\star$ is the surface density of all the stellar mass that has been formed and not the one we observe today, that is corrected according to the mass loss due to stellar evolution. We take $t_{univ}=13.8$ Gyr. Doing so (see Figure \ref{fig:cosmic_SFRdens}), we find that for spirals, both the profiles of the log $\langle \Sigma_{\text{SFR}}(R)\rangle _{cosmic}$ and the log $\Sigma_{\text{SFR}}$ have similar values above the reference line, with the latter being shallower. We note, that as we move to later types, the intersection between the two profiles happens closer to the galactic centre, indicating that a larger portion of the disc is forming stars at a higher rate than the mean rate across cosmic time. On the other side, early types are clearly quenched: the values for log $\langle  \Sigma_{\text{SFR}}(R)\rangle _{cosmic}$ at R = 1 HLR are $1.50\pm0.06$ and $1.77\pm0.04$ [M$_\odot$/yr/pc$^2]$ for E and S0, and the $\Sigma_{\text{SFR}}(R)$ at the same distance is much lower for these types, $0.20\pm1.23$ and $0.26\pm0.25$ [M$_\odot$/yr/pc$^2]$, respectively. These results are similar to those in \citet{Gonzalez-Delgado2016}.

\subsection{sSFR} \label{subsection: results sSFR}

\begin{figure}
\centering
\includegraphics[width=0.45\textwidth]{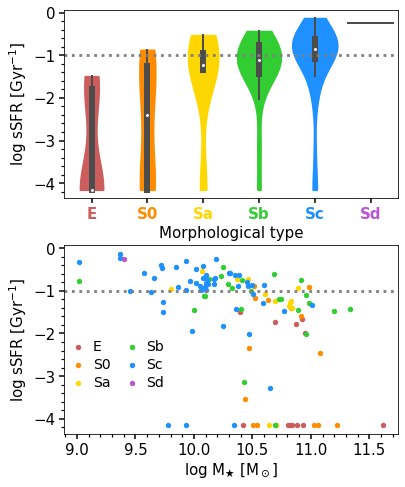}
\caption{Global specific SFR. The galaxies with null sSFR were given the value sSFR = 10$^{-4.15}$Gyr$^{-1}$ (the value of the minimum minus the standard deviation) so that they could appear in the plots. The dotted line marks the limit between star-forming and quenched galaxies, adopted in \citet{Peng2010}. \textit{Top}: Distribution of the total sSFR for each type (violin plot as in Figure \ref{fig:mass_morph}). \textit{Bottom}: sSFR versus stellar mass.}
\label{fig:integ_sSFR}
\end{figure}

The sSFR is a measurement of the relative rate at which galaxies are forming stars in the present, with respect to the past average rate. The distribution of the total sSFR for each morphological type is shown in the top panel of Figure \ref{fig:integ_sSFR}. Galaxies with null sSFR were attributed the value log sSFR [Gyr$^{-1}]=-4.15$ to maintain their presence in the plots, the minimum value minus the standard deviation, as we did for the SFR. As it happened for the SFR (Figure \ref{fig:integ_SFR}), the sSFR increases with later morphological types. We find the largest dispersion at early types, which agrees with previous studies \citep{Schiminovich2007, Salim2007}.

The bottom panel shows the relation between the sSFR and the stellar mass. Because the SFMS is a sublinear regression with $M_\star$ (SFR $\sim \text{M}_\star^\alpha$, with $\alpha<1$), we expect the sSFR to decrease with stellar mass: sSFR $\sim \text{M}_\star^{\alpha-1}$. This means that we can 'translate' the SFMS in terms of the sSFR and get a sSFR-M$_\star$ relation. One can see this decreasing sequence in our data. Early-type galaxies, which are mainly out of the SFMS, will also be out of this relation.

\citet{Peng2010} defined a threshold between star-forming galaxies and quenched systems at sSFR = 0.1 Gyr$^{-1}$. This value represents the sSFR that a galaxy should have to grow its current stellar mass at the present rate during a Hubble time of 10 Gyr. Comparing with the global values we obtained, we can check that ETG (Es and S0s) are very far below this value, Sas and Sbs are closer to it, and Scs and the Sd surpass it.

\begin{figure}
\centering
\includegraphics[width=0.45\textwidth]{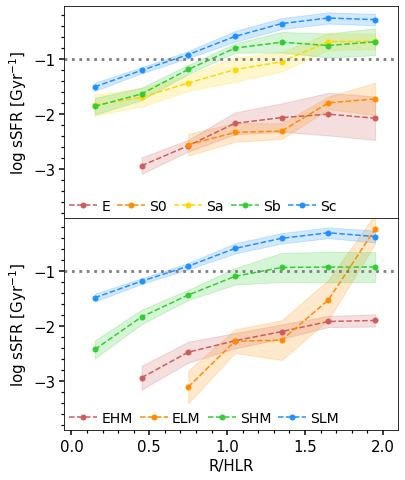}
\caption{Radial profiles of the intensity of the specific SFR. The dotted line marks the limit between star-forming and quenched regions, adopted in \citet{Peng2010}. \textit{Top}: Stacked radial profiles by morphological type. \textit{Bottom}: Stacked radial profiles in bins of morphology and mass. See Figure \ref{fig:prof_Mdens} for more details.}
\label{fig:prof_sSFR}
\end{figure}

The stacked radial profiles of the sSFR are shown in Figure \ref{fig:prof_sSFR}. As we did with the profiles of the SFR density, we take every galaxy into account, including those with global SFR = 0. All the galaxies show a radial structure of the sSFR that increases outwards. The result resembles the one from the intensity of the SFR: early types have almost no sSFR, and show lower values than spiral types, well below the 0.1 Gyr$^{-1}$ threshold. It happens in both panels, where the difference between these two categories is very noticeable. Only the inner part (R < 1 HLR) of the spirals has lower values than the reference, probably signaling the bulge-disc transition. We find that the sSFR profiles scale with morphology (top panel), with values at R = 1 HLR of $-2.17\pm0.20$, $-2.34\pm0.17$, $-1.20\pm0.22$, $-0.81\pm0.08$, and $-0.60\pm0.10$ for E, S0, Sa, Sb, and Sc, all in units of Gyr$^{-1}$ in logarithmic scale.

Similarly, the sSFR radial structure also scales with the total stellar mass (bottom panel of Figure \ref{fig:prof_sSFR}): less massive spirals have higher sSFR than more massive ones at every radial distance. However, early types show similar sSFR values independently of their M$_\star$, except for the outer parts, where low mass early type galaxies seem to have an increase in the sSFR. However, this effect might be caused by the low number of galaxies in this category, accentuated by the higher percentage of quenched ETG.

\begin{table*}
\centering
\begin{tabular}{cccccc}
HType & log M$_\star$ [M$_\odot$] & log $\Sigma_\star$ [M$_\odot$/pc$^2$] & $\langle \text{log t} \rangle _L$ [yr] & log SFR [M$_\odot$/yr] & log sSFR [Gyr$^{-1}$] \\ \hline
E & $10.84\pm0.28$ & $2.33\pm0.13$ & $9.68\pm0.22$ & $-$ & $-$ \\
S0 & $10.73\pm0.27$ & $2.43\pm0.18$ & $9.61\pm0.28$ & $-$ & $-$ \\
Sa & $10.69\pm0.31$ & $2.44\pm0.19$ & $9.24\pm0.33$ & $0.53\pm0.27$ & $-1.08\pm0.29$ \\
Sb & $10.46\pm0.48$ & $2.26\pm0.28$ & $9.16\pm0.35$ & $0.49\pm0.65$ & $-1.12\pm0.57$ \\
Sc & $10.10\pm0.38$ & $1.87\pm0.27$ & $8.70\pm0.34$ & $0.29\pm0.51$ & $-0.83\pm0.54$ \\
Sd & 9.41 & 1.46 & 8.23 & 0.16 & $-0.24$ \\
\end{tabular}
\caption{Global median values and standard deviation for each morphological type of the derived stellar properties. The last row refers to the value of the single Sd galaxy in the sample. The log SFR and log sSFR were calculated without taking into account the galaxies with SFR = 0.}
\label{tab:median_global_results}
\end{table*}

\section{Comparison with galaxies in filaments and walls} \label{Section: comp CALIFA}

The main purpose of this paper is to achieve insight into the role that the low density environment plays in the global and spatially resolved stellar properties of galaxies. So far, we have obtained that $\Sigma_\star$, $\langle \text{log t}\rangle _L$, $\Sigma_{\text{SFR}}$, and sSFR show similar radial distributions to those of galaxies located in other environments (like the galaxies in the CALIFA survey). Here, we discuss in detail the role of voids in setting different properties in galaxies by doing a fair comparison with galaxies that are located in higher-density environments, such as filaments and walls. 

\subsection{Control sample selection} \label{Subsection: control sample selection}

To build the control sample, we select galaxies in filaments and walls from the CALIFA survey \citep{Sanchez2012}, that were observed with the same instrument as CAVITY and had been fitted following the same methodology (same algorithm and same SSPs). We start from the third Data Release \citep{Sanchez2016}, which consists of 646 low-resolution (V500) datacubes. We then remove those that belong to the extended sample, that targets more 'unusual' galaxies (ultra-compact, mergers, etc.) and those that do not have a determined morphological type. We end up with 514 galaxies.

Other studies about galaxies in voids choose the control sample as a general selection of any galaxy that does not belong to a void. However, we find this decision misleading: many works have already shown the effect that belonging to a cluster can have on the evolution and stellar properties of a galaxy \citep{Balogh2004, Blanton2005, McNab2021, Sobral2022, Gonzalez-Delgado2022, Rodriguez-Martin2022}. The differences between galaxies in voids and galaxies in a general sample of non-void galaxies will probably be dominated by the high-density environment galaxies. This effect is already widely known and would give us no information about the impact of the void environment on galaxy evolution. Since we want to avoid this, we exclude galaxies in the highest-density environments from our control sample, leaving only those in filaments and walls, where the matter density is close to the mean density of the Universe.

To make sure that our CALIFA subsample consists of galaxies that belong to filaments and walls, we cross-check it with two catalogues: \citet{Tempel2017} for galaxies in groups and clusters (defined as having more than 30 members, \citealt{Abell1989}, same criteria as in \citealt{Dominguez-gomez2022, Dominguez-Gomez2023a, Dominguez-Gomez2023b}) and \citet{Pan2012} for voids, taking out those with an effective radius fraction $\leq$ 1. This parameter represents the distance to the centre of the void normalised by the void radius (considered as spheres as a first approximation), so values above 1 are given to galaxies that inhabit the edge of the voids. After this second removal, we are left with 410 galaxies in filaments and walls.

This method can serve as a first approximation: filaments are complex structures cover a vast range in density (5 orders of magnitude), ranging from the dense and heavy filamentary inflow arms to clusters, to underdense tendrils in the interior of voids \citep{Cautun2014}. Additionally, groups of galaxies are found inside filaments, which result in an inhomogeneous mass distribution. Several cosmic web identification formalisms are focused on the detection and the study of the substructures in filaments (e.g. MMF/NEXUS, \citealt{Cautun2013}, Disperse \citealt{Sousbie2011}, Bisous, \citealt{Tempel2016}, Vweb, \citealt{Hoffman2012}). However, for practical purposes, we consider our approximation to be reasonable enough for this work.

Since 
the fraction of red galaxies is strongly mass dependent \citep{Kauffmann2003, Baldry2006}, we need to make sure that both samples are well matched in mass. The same is the case for morphological type, as both attributes are very relevant to determine the evolutionary properties of galaxies \citep{Perez2013, Gonzalez-Delgado2017, Lopez-Fernandez2018}. For this reason, the last step is to match the control sample as close as we can in total stellar mass and morphology to the void sample. For each CAVITY galaxy, we chose the CALIFA one with the same morphological type that has the closest total stellar mass. This way we build a pair sample for each void galaxy, checking that they do not repeat. This procedure has been followed to obtain samples of twin galaxies in previous works (e.g. \citealt{delMoral-Castro2019, delMoral-Castro2020}).

The morphological classification of CALIFA galaxies was carried out through a visual assessment by a group of members of the collaboration, as detailed in \citet{Walcher2014}. We did not use the classification from \citet{Dominguez-Sanchez2018}, the one used for the CAVITY galaxies, as only 12 of our selected galaxies were in the catalogue. However, we consider it to be compatible with the one we use, given that their Deep Learning algorithms are trained with data that is classified by visual inspection.

\begin{figure}
\centering
\includegraphics[width=0.45\textwidth]{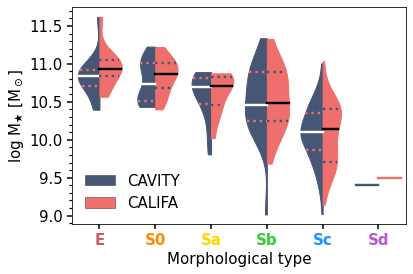}
\caption{Distribution of the total stellar mass by morphological types, for CAVITY void galaxies (in dark blue) and for CALIFA galaxies in filament or walls (in red). The coloured area determines the density distribution (Gaussian kernel), the central black or white continuous lines mark the median values of each distribution, and the dotted lines are the first and third quartiles. The edges of the distributions are cut to define the minimum and maximum values. The Sd category only shows a line with the value for its single galaxy.}
\label{fig:dist_logM_morph_CALIFA}
\end{figure}

The distribution of the total stellar mass for each morphological type for the final control sample and the CAVITY sample is shown in Figure \ref{fig:dist_logM_morph_CALIFA}. For most morphological types, both distributions match reasonably well, but for early types (E and S0) there is a small mismatch towards more massive galaxies in the CALIFA sample. The median values and dispersion of the logarithm of the stellar mass (in solar masses) for galaxies in voids and galaxies in filaments and walls are $10.84\pm0.30$ and $10.93\pm0.26$ for E, and $10.73\pm0.28$ and $10.87\pm0.24$ for S0, so we do not find this difference to be significant enough. The only way to solve this issue was to remove void galaxies for the comparison, which would reduce a not-so-abundant sample and leave us with worse statistics, so we decided to leave it this way.

To quantitatively assess the similarity of the samples, we employ a two-sample Kolmogorov-Smirnov test (K-S test). This statistical test is designed to compare two independent samples, determining whether both are drawn from the same distribution. The test outcome is expressed through a parameter known as the p-value, which tells the probability for this to happen. A threshold p-value is chosen in order to decide whether two samples are different, which usually is p < 0.05 (a confidence level of 95\%). The p-values we obtain for each morphological type, from E to Sc, are 0.30, 0.34, 1.00, 1.00, and 0.75. Taking all the morphological types together to compare the overall distribution we obtain a p-value of 0.88. All of these values are higher than the threshold, so we conclude that the two samples are similar enough in mass and morphology.

\subsection{Global stellar population properties}

\begin{figure*}
\centering
\includegraphics[width=\textwidth]{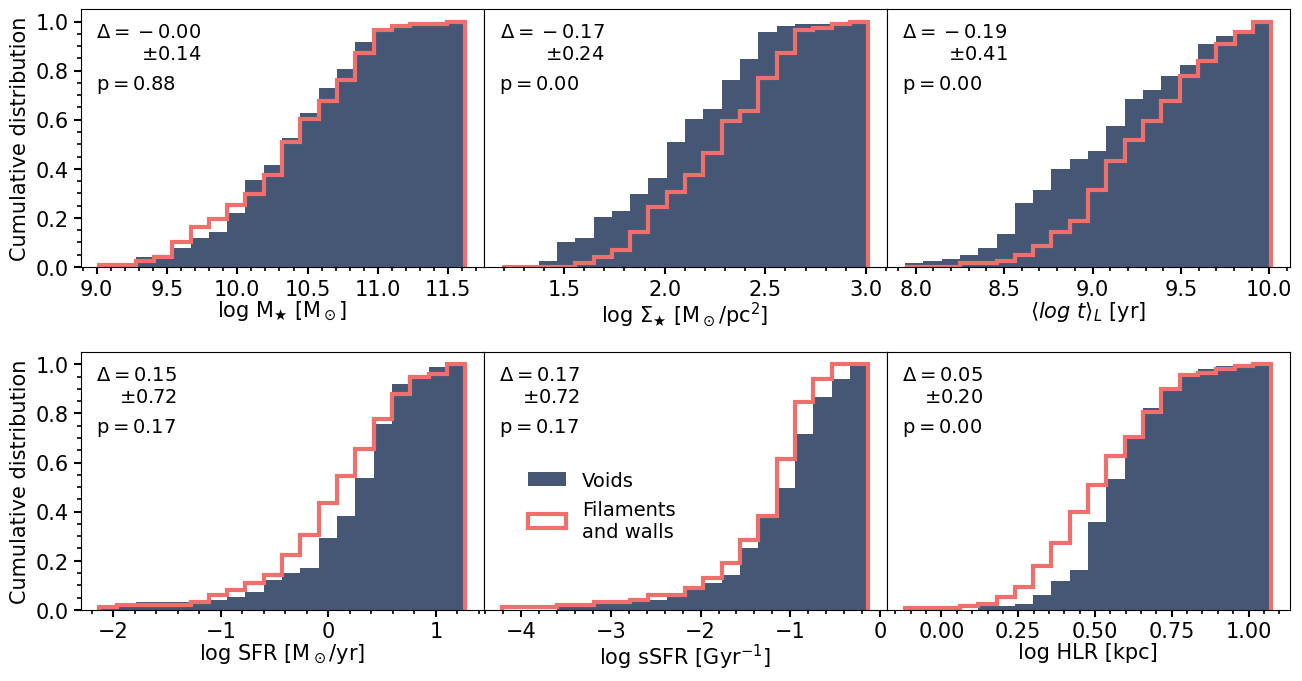}
\caption{Cumulative histograms of the global stellar population properties and HLR for voids (in dark blue) and filaments and walls (overlined in red). The properties plotted are, from left to right, stellar mass, stellar mass density, light-weighted age, SFR, sSFR and half-light radius. The median and the standard deviation of the subtraction between voids and filaments and walls ($\Delta=\Delta_\text{void-fil}$) is shown in the top left or right corner of each panel. Below, it is shown the p-value resultant of a two-sample K-S test applied to the distributions of the two samples.}
\label{fig:integrated_prop_hist_comp}
\end{figure*}

First, we compare the global properties of both environment samples through their cumulative histograms (Figure \ref{fig:integrated_prop_hist_comp}). We do the subtraction between the values of each pair of twin galaxies ($\Delta=\Delta_\text{void-fil}$) for each property, and then obtain the median and standard deviation of the distributions, shown in the top corner of each panel. Additionally, we apply to the distributions of each stellar property a K-S test. The p-value for each property is shown under the $\Delta$ in each panel. Given the similarity in the mass distributions, seen in the top left panel, with a median deviation of $0.00\pm0.14$ [M$_\odot]$ and a p-value of 0.88, we decided to interpret the differences between the rest of the properties through the histograms rather than using scatter plots against the stellar mass. Additionally, we calculated the median difference for each morphological type, shown in Table \ref{tab:diff_global_prop}. The uncertainty is the standard deviation of the distributions, nonexistent for the only Sd galaxy in the last row.

\begin{table*}
\centering
\begin{tabular}{ccccccc}
HType & $\Delta$log M$_\star$ [M$_\odot$] & $\Delta$log $\Sigma_\star$ [M$_\odot$/pc$^2$] & $\Delta$$\langle \text{log t} \rangle _L$ [yr] & $\Delta$log SFR [M$_\odot$/yr] & $\Delta$log sSFR [Gyr$^{-1}$] & $\Delta$log HLR [kpc] \\ \hline
E & $-0.06\pm0.18$ & $-0.24\pm0.20$ & $-0.07\pm0.29$ & $-0.15\pm1.20$ (3) & $0.54\pm1.15$ & $0.14\pm0.13$ \\
S0 & $-0.00\pm0.13$ & $-0.11\pm0.25$ & $0.03\pm0.31$ & $1.51\pm0.03$ (2) & $1.52\pm0.02$ & $0.06\pm0.16$ \\
Sa & $-0.01\pm0.09$ & $-0.19\pm0.20$ & $-0.26\pm0.36$ & $0.40\pm0.44$ (10) & $0.44\pm0.44$ & $0.08\pm0.21$ \\
Sb & $-0.00\pm0.14$ & $-0.17\pm0.26$ & $-0.09\pm0.48$ & $0.14\pm0.81$ (23) & $0.14\pm0.82$ & $-0.03\pm0.18$ \\
Sc & $0.00\pm0.13$ & $-0.18\pm0.25$ & $-0.33\pm0.41$ & $0.13\pm0.60$ (50) & $0.13\pm0.58$ & $0.02\pm0.21$ \\
Sd & -0.09 & -0.37 & -0.33 & 0.15 (1) & 0.24 & 0.15 \\
\end{tabular}
\caption{Median difference of the global stellar population properties and HLR between galaxies in the two samples (voids - filaments and walls), for each morphological type. 
The uncertainty is given by the standard deviation of the distribution. The last row belongs to the difference in the values of the single Sd galaxy in each sample. The values between brackets in the $\Delta$log SFR column refer to the number of pairs taken into account to calculate the difference in log SFR and log sSFR (those in which none of the values were masked).}
\label{tab:diff_global_prop}
\end{table*}

The distribution of the stellar mass surface density (upper middle panel)
shows a shift of $-0.17\pm0.24$ [M$_\odot$/pc$^2]$ towards lower densities in voids. This difference is present for all morphological types, but more prominent for elliptical galaxies, where the median value of log $\Sigma_\star$ [M$_\odot/$pc$^2]$ for galaxies in voids is $2.33\pm0.13$, and $2.60\pm0.19$ for galaxies in filaments and walls. A p-value of 0 tells us about the dissimilarity of the distributions.

The upper right panel refers to the distribution in $\langle \text{log t}\rangle _L$, where the difference of $-0.19\pm0.41$ [yr] and a p-value of 0 suggests that galaxies in voids are younger compared to those in filaments and walls. This age difference is observed across all morphological types except for S0s and is more pronounced for Scs, followed by Sas.

The SFR (lower left panel) seems to be higher in voids than in filaments and walls. 
Here, we did not assign any value to the galaxies with null SFR, as we did in Section \ref{subsection: results SFR}, so they are not taken into account in when doing the histogram or calculating the medians and standard deviations. We find a median shift of $0.15\pm0.72$ [M$_\odot/$yr], which is expected due to the correlation between the ages of the stellar populations and the SFR: the younger the age, the higher the SFR \citep{Asari2007}. \citet{Duarte-Puertas2022} noted, using galaxies from the SDSS, that this correlation stands for star-forming galaxies at fixed stellar mass.
We find that the difference is higher for S0s, followed by Sas: those that are in transition between spiral and quiescent. We must take into account that when doing this calculation, the difference of pairs of twin galaxies in which one of the members has SFR = 0 is masked, so the morphological types with a larger amount of quiescent galaxies end up with fewer statistics. The number of pairs that are taken into account when calculating the median and standard deviation of the log SFR and log sSFR is 3, 2, 10, 23, and 50, starting and E and ending at Sc.

The lower middle panel shows similar results for the sSFR to those of the SFR (lower left panel). This is to be expected, given that the mass makes no difference because of the way the control sample is chosen. The median shift is $0.17\pm0.72$ [Gyr$^{-1}]$. When looking at the median differences in galaxies of the same type, we find that they are larger for ETG and early spirals. Opposite to what happened for the log SFR, this time elliptical galaxies show a large positive difference (more sSFR in voids), due to the shift towards lower stellar mass in voids. However, we must remember that when talking about the SFR and sSFR, the results for Es and S0s may not be reliable, as we do not have a large number of galaxies. The log SFR and log sSFR distributions return the highest p-value of all properties, apart from the stellar mass. These identical values, p = 0.17, are above 0.05, which is usually taken as limit value. This means that the differences between samples are not statistically relevant enough to draw strong conclusions. If we check the p-value when comparing the distributions for each morphological type, we find that only in the case of the Sas p is found to be lower than 0.05, for both the SFR and sSFR.

Although the number of quiescent galaxies (those with SFR = 0) in voids and in filaments and walls in our sample was found to be the same, if we check how this amount changes with morphological type, we see that the distribution is not the same for both samples. For ellipticals, we do not find that much difference (62\% in voids versus 69\% in filaments and walls). Some spiral galaxies are quiescent in voids that do not find a matching behaviour in filaments, but they only represent 0.6\% of the total spiral galaxies. The largest difference appears for lenticular galaxies, where 64\% of filament and wall galaxies are quenched, but only 36\% are in voids. This could mean a slower transition between star-forming and quiescent: galaxies in voids take longer to halt their star formation for a similar morphological type, and between the ETG that are still forming stars, those that are in voids are doing so at a higher rate, being further from reaching a quenched state.

\subsection{Half-light radius}

For the calculation of the radial profiles of the galaxies in the control sample, we obtain the HLR in the same way as we did for the void galaxies datacubes (in Section \ref{subsection: HLR}). The lower right panel in Figure \ref{fig:integrated_prop_hist_comp} shows the distributions for both samples. We find that in general, galaxies in voids are slightly more extended than galaxies in filaments and walls, with a median difference of $0.05\pm0.20$ in log HLR [kpc]. The difference between the distributions observed in the histograms concurs with a resultant p-value of 0. When checking by morphological types, we see that the biggest difference appears for ellipticals, and that the tendencies mainly match what was obtained for the stellar mass surface density. This is to be expected, given that the difference in the mass is $\sim$ 0, and that the stellar mass density can be approximated as $\Sigma_\star=\text{M}_\star/\text{area}\sim \text{M}_\star/4\pi(\text{HLR})^2$.

\begin{figure}
\centering
\includegraphics[width=0.45\textwidth]{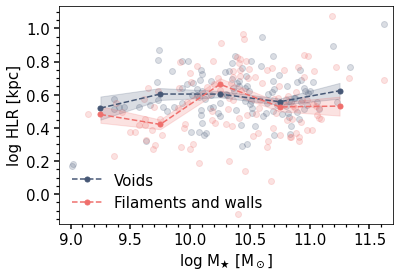}
\caption{Half-light radius versus total stellar mass for the galaxies in voids (in dark blue) and the galaxies in filaments and walls (red). The darker points connected by dashed lines refer to the median values in bins of stellar mass. The shaded area around them marks the error of the mean (the standard deviation divided by the square root of the number of points in each bin).}
\label{fig:hlr_mass_comp}
\end{figure}

The size of a galaxy shows a dependence on the stellar mass, also called the mass-size relation, and it is thought to be another indicator of evolution in a galaxy: larger galaxies are usually more massive, probably as a result of mergers \citep{Robertson2006, Naab2009}. The size of void galaxies was addressed in \citet{Porter2023}, where they found no difference between the mass-size relation in voids and in filament and wall galaxies. However, for constructing their control sample they did not only match the morphological type and the mass, but also the sSFR, so our results are not totally comparable. Our results (in Figure \ref{fig:hlr_mass_comp}) indicate that the size of void galaxies and galaxies in filaments and walls is similar, except for those with stellar masses between $10^{9.5}$ and $10^{10}$ M$_\odot$, where the HLR is larger in voids. 

\subsection{Radial profiles}

We compare the radial structure of the stellar population properties of galaxies in voids and in filaments and walls, to determine if the differences identified in the global properties are occurring everywhere in the galaxies, or instead, they are specifically associated with the inner regions (R < 1 HLR) or outskirts (R > 1 HLR) of the galaxies. While the light from the central part of S0s and spirals is probably more associated with the bulge, the light from regions located at distances larger than 1 HLR comes from the discs. Thus, finding out where the variations come from can give us clues about different formation scenarios and/or evolution of the bulges and discs of galaxies in low-density environments.

\begin{figure*}
\centering
\includegraphics[width=\textwidth]{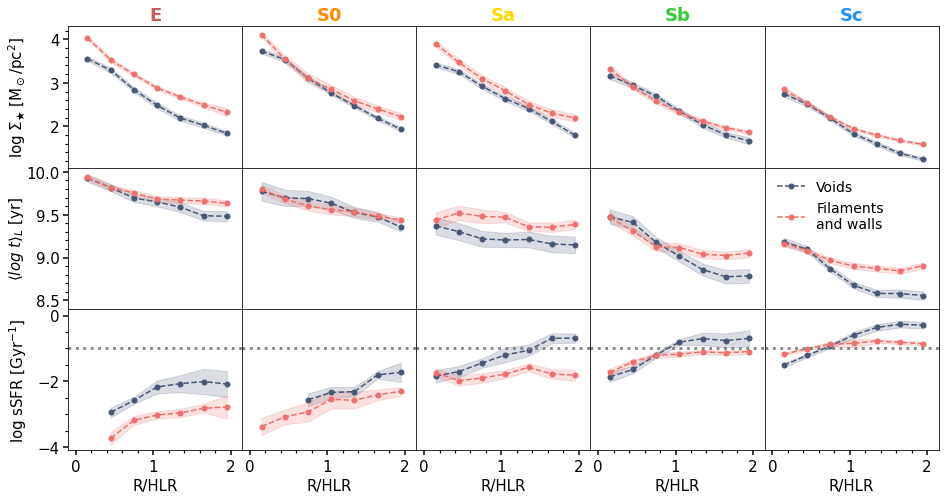}
\caption{Radial profiles of the stellar population properties for galaxies in voids and in filaments and walls, stacked by morphological types in different panels. From top to bottom, stellar mass density, age, and sSFR. The dotted grey line in the lower row of panels refers to the limit between star-forming and quiescent from \citet{Peng2010}.}
\label{fig:prof_comp}
\end{figure*}
    
Figure \ref{fig:prof_comp} shows the radial profiles of the properties we are analysing. In each panel, we represent the stacked profiles of each morphological type of the void sample and the filament and wall sample. Like before, we decided not to plot the Sd, as we only have one.

The radial profiles of the stellar mass density (upper panels) indicate that the largest difference occurs for elliptical galaxies, with $\Delta_{\text{void-fil}}\text{log }\Sigma_\star(\text{1HLR})=-0.32\pm0.10$ [M$_\odot$/pc$^2$]. This shift is present at any galactocentric distance. There is also a mismatch between the profiles in the outer parts ($\text{R}>1\,\text{HLR}$) of spiral galaxies and S0s, and in the bulges of S0s and Sas ($\text{R}<0.3\,\text{HLR}$), where void galaxies are less dense (have lower $\Sigma_\star$).

In the case of the light-weighted age, on the panels of the second row, we find the opposite: there is no significant difference in the earlier types, but in the spirals, where we see a mismatch in the outer parts. We find the biggest difference from 0.5 HLR outwards, outside of what could be the bulge of the spiral types. The discs of spiral void galaxies seem to be less evolved, with more pronounced gradients than those from galaxies in filaments and walls. In the Sas, we see differences not only in the outer parts but at every galactocentric radius except for the central region. 

The last panels refer to the sSFR. They follow the same tendencies as the age.

Early-type galaxies in voids tend to have higher sSFR than galaxies in filaments and walls. However, in both cases, these values are far below the reference value 0.1 Gyr$^{-1}$, indicating that the star formation is quenched at every galactocentric distance. The opposite happens in late-type spirals (Sbs and Scs), which are very close to or above this limit, implying that they are actively forming stars in their discs. It is there (at R > 1HLR) where the differences emerge, and we see that the discs of void galaxies reach higher values of sSFR, while their inner parts have similar values to those from galaxies in filaments and walls. In the case of early-type spirals, the differences between environments are significant; while the outer parts of Sas in voids are still in the SFMS as other spirals, their twins in filaments and walls are already outside this relation with sSFR < 0.1 Gyr$^{-1}$. 

This result, along with those from the age, points to a difference in the discs of spiral galaxies between galaxies in voids and galaxies in filaments and walls: those in voids are less evolved and more star-forming. It matches with what one could expect, as interactions or ram-pressure stripping have a larger effect on the outer parts of galaxies, heating or removing the gas and limiting the ability of the galaxy to form stars. This stronger steepness in the SFR profiles for galaxies in denser environments (and weaker in sSFR profiles) was noted in \citet{Schaefer2017} and \citet{Medling2018}.

In the case of the ellipticals, the increase in the SFR does not match with the age of the galaxies but with their density. The physical mechanisms behind this phenomenon could be a recent morphological transition, that has left not enough time for the galaxy to exhaust its gas \citep{Haines2015}, or cold gas accretion from the low-density environment \citep{Thomas2010}. This result agrees with what was stated in \citet{Florez2021}: that early-type galaxies in voids are bluer and more star-forming than galaxies of the same type and mass range in filaments and walls. 

As expected, the spatially resolved results coincide with the tendencies observed through the global properties.

\section{Discussion} \label{Section: Discussion}

In this section, we try to understand the physical mechanisms that drive galaxy evolution in different environments, in relation with our results and previous works. We discuss about star formation histories, quenching mechanisms, in situ star formation, gas accretion, mass assembly, and mergers.

\subsection{Evolution of galaxies in voids}

Some works in the literature have already addressed this topic, through different evolutionary tracers. Evidence of lower gas oxygen abundances in void galaxies in \citet{Pustilnik2011, Pustilnik2016, Pustilnik2024, Kniazev2018} is explained as a result from slower evolution in voids than in denser environments. They also found a population of possible Very Young Galaxies among the least luminous void objects \citep{Pustilnik2020, Pustilnik2021}.

On the same line, one of the most recent results regarding the differences in the SFHs between different environments comes from \citet{Dominguez-Gomez2023b}.
To do so, they use spectra from the SDSS of CAVITY galaxies, as well as of galaxies in other environments: filaments and walls, and clusters. They also split the sample in two, depending on their SFH type: short timescale SFH (ST-SFH), which contains galaxies that have formed most of their mass at the earliest time, and long timescale (LT-SFH), for those that have had more homogeneous star formation over time. Given the nature of SDSS data, the properties obtained refer to the central parts of the galaxies. Through their analysis, they concluded that galaxies in voids evolve at a slower pace than galaxies in more dense environments.  This result is more significant for galaxies with LT-SFH and low stellar mass, $\text{log M}_\star[\text{M}_\odot]\sim10$.

Other work on the matter is that of \citet{Parente2023}. They used semi-analytic galaxy formation model L-Galaxies \citep{Henriques2020}, on top of the Millenium simulation \citep{Springel2005}, as well as observational stellar properties of galaxies obtained from the MPA-JHU catalogue \citep{Kauffmann2003}. Galaxies in both samples were classified according to their environment, in voids, walls, filaments, or clusters (nodes). Then, they calculated the sSFR and mass-weighted stellar ages as a function of the mass of the galaxy. 
The mean age of a galaxy can be interpreted as a simple way to compress its SFH. Usually, the measure chosen for this purpose is the stellar mass-weighted age (instead of the light-weighted age), as it is believed to be a more robust SFH tracer \citep{Gonzalez-Delgado2015}. They obtain very similar results from both simulated and observed data: there is a larger difference between environments at lower stellar masses (with younger galaxies and higher sSFR in environments of less density), but these differences decrease as mass increases until a critical mass of $\sim 10^{10.8}\text{ M}_\odot$. After reaching this limit, galaxies in all environments share the same values due to the prevalence of internal processes that are responsible for halting the star formation even in isolated systems. We will discuss these processes associated with mass quenching in Section \ref{Subsection: quenching}.

\begin{figure}
\centering
\includegraphics[width=0.45\textwidth]{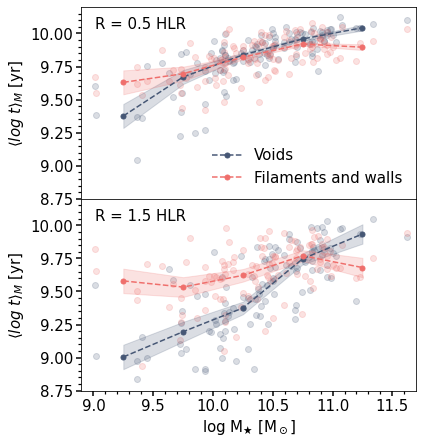}
\caption{Mass-weighted age versus stellar mass for the galaxies in voids and the galaxies in filaments and walls (see Figure \ref{fig:hlr_mass_comp} for more details). \textit{Top}: Age at a R = 0.5 HLR, representative of the inner part of the galaxies. \textit{Bottom}: Age at a R = 1.5 HLR, representative of the outer part of the galaxies.}
\label{fig:age_zones_mass_comp}
\end{figure}

In order to compare with these studies, we check how the age of galaxies in our two environments changes as a function of their mass. Taking advantage of the fact that we have spatially resolved information, we obtain the ages in the inner parts of the galaxies (at R = 0.5 HLR) and in their outer parts (at R = 1.5 HLR, see Figure \ref{fig:age_zones_mass_comp}). It shows that galaxies in voids have lower values of mass-weighted ages. This is a consequence of slower (or more homogeneous over time) SFHs, which is consistent with the results in \citet{Dominguez-Gomez2023b}.  Further, the $\langle\text{log t}\rangle_M$ measured at R= 1.5 HLR (bottom panel in Figure \ref{fig:age_zones_mass_comp}),  is very similar to the results in \citet{Parente2023}. The threshold mass they define for efficient mass quenching coincides with the mass bin between 10$^{10.5}$ and 10$^{11}$ M$_\odot$, where the lines corresponding to voids and to filaments and walls converge. In the last mass bin, the trends are inverted, but one must note that the number of points is fairly low.

\subsection{The quenching of star formation in voids} \label{Subsection: quenching}

The rapid transition between galaxy types, from bluer and star-forming, to redder and passive, is called quenching. This phenomenon can be driven by different physical mechanisms, that are mainly separated into two categories: mass quenching and environmental quenching \citep{Peng2010}. The mass quenching is due to internal processes which only concerns the galaxy on its own and its mass \citep{Peng2012, Arcila-Osejo2019, Contini2020, Guo2021}, as AGN feedback, which can heat up infalling gas and stop the formation of new stars \citep{DiMatteo2005, Fabian2012, McNamara2012}. Environmental quenching, on the other hand, comes from effects external to the galaxy, such as interaction with other galaxies or with the intergalactic medium. High density environment can sometimes cause an enhancement in the SFR, but its effect on the long term is the shutting down of the star formation because of the heating or removal of the gas in the galaxies \citep{Alatalo2015, Davies2015, Lisenfeld2017, Joshi2019}.

Quenching is caused by a combination of these effects. It is known that the fraction of galaxies that have undergone this transformation is higher in environments of high density \citep{George2011}, especially near the centres of clusters \citep{Donnari2021}. Furthermore, the fraction of quenched galaxies in groups increases with galaxy stellar mass, but is always higher in groups than in the field \citep{Gonzalez-Delgado2022}. We want to test if a similar difference is also found for the environments of least number-density with respect to filaments and walls.

This has already been assessed in \citet{Rosas-Guevara2022}, using simulated data from the EAGLE project \citep{Schaye2015, Crain2015} and the \citet{Paillas2017} void catalogue. They separate the sample into galaxies in filaments (or skeleton), walls, inner part of the voids (inner void) and in their outer part (outer void). 
They find differences between the fraction of quiescent galaxies in the filaments and walls categories with respect those in the inner and outer voids, which depend on the galaxy stellar mass. They point to a lower fraction of quenched galaxies in voids, except for galaxies of $\sim10^{10}\text{ M}_\odot$,  where the fraction is slightly higher in voids.

To compare with their result, we also separate the galaxies between star-forming and quiescent, applying the definition used in \citet{Bluck2016}. They computed the difference between the SFR of a galaxy and what would correspond in the SFMS at their mass ($\Delta\text{SFR}$), and found a bimodal distribution, from which they define galaxies to be quenched if their $\Delta\text{SFR}$ is more than 1 dex below to the SFMS. 

\begin{figure}
\centering
\includegraphics[width=0.45\textwidth]{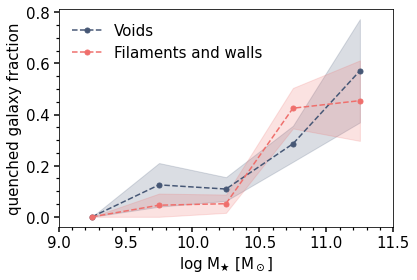}
\caption{Percentage of quenched galaxies per mass bin for the two environments, calculated as those with a deviation from the SFMS $\Delta\text{SFR}<-1$ \citep{Bluck2016}. The shaded area corresponds to jackknife errors.}
\label{fig:quenched_fraction}
\end{figure}

In our results (Figure \ref{fig:quenched_fraction}), we do not see any significant difference between the two environments, when the errors that are estimated through the jackknife technique, are taken into account.
But we must remember that our sample is not only smaller, but also matched in morphology and stellar mass, which could affect the results. It would be interesting to see if this fraction changes as the distance to the centre of the void increases, as it happens in clusters. We plan to do this calculation in a future work, as well as checking the rest of the stellar population properties, once we have more data. 

\subsection{In situ star formation and gas accretion mode in void galaxies}

The sSFR tells us about the current evolutionary stage of a galaxy, and measures how actively is it now forming stars with respect to its past history. Complementary to the results retrieved from observational data of the galaxies' current stage, simulations can give us a much more detailed understanding of the star formation and gas accretion histories that lead to what we observe today.

In \citet{Beygu2016A}, they associate the higher values of the sSFR in voids to a more steady inflow of gas (cold accretion). This process was first discussed by \citet{Dekel2006}, who proposed that the star-forming galaxies with $\text{M}_\star<3\times10^{10}\text{M}_\odot$ reside in haloes whose mass is below a critical shock-heating mass, and where the disc is built by cold flows and might generate early starbursts. The dominance of cold mode accretion in low density regions was also suggested in \citet{Keres2005}, using simulations.

This agrees with what was found in \citet{Galarraga-Espinosa2023}, where they studied through data from the TNG50-1 simulation (the highest resolution box of the IllustrisTNG simulation, \citealt{Pillepich2018, Nelson2019}) and the DisPerSE filament finder algorithm \citep{Sousbie2011} the relation between the SFR and the galaxy connectivity: the amount of streams of gas that are connected to a galaxy. They found that a higher number of streams connected to a galaxy results in a higher sSFR, and that this happens more commonly as the density of the LSS environment decreases, given that tidal forces, which could disrupt the gas streams, are weaker.

Hints of recent cold gas accretion from filamentary substructures within the voids into galaxies have been observed in previous works, as it is the case of VGS12 \citep{Stanonik2009}, VGS31 \citep{Beygu2013}, VGS38 \citep{Kreckel2011}, and UGC3672 \citep{Chengalur2017}. Observational evidence of this substructure has been recently found in the Local Void by \citet{Kurapati2024}.

Differences in the star formation properties may also be due to changes in the relation between baryonic and dark matter in different environments. Works based on simulations \citep{Alfaro2020, Habouzit2020, Rosas-Guevara2022, Rodriguez-Medrano2024}, find that for a fixed stellar mass, dark matter haloes are more massive in voids than in denser environments. This means that for a fixed stellar mass, the gravitational potential well in which galaxies evolve is higher in voids, which can affect their properties. It has been shown that the mass of the haloes affect the SFHs of the galaxies that inhabit them \citep{Scholz-Diaz2023}, and that higher halo mass-to-stellar mass ratios result in younger stellar ages \citep{Scholz-Diaz2022}, which is consistent with our results.

Previous works draw different conclusions when assessing the differences in the global sSFR. Some of them, as \citet{Rojas2005}, \citet{Beygu2016A}, \citet{Moorman2016} and \citet{Florez2021}, obtain higher sSFR in void galaxies, through different methodologies. However, other studies find that galaxies in voids and filaments exhibit similar sSFR \citep{Ricciardelli2014, Patiri2006, Dominguez-gomez2022, Rosas-Guevara2022}. 

In this work, we add to the discussion a new way to look at the results: through spatially resolved data. Where we find the strongest differences between environments is always on the outer parts of galaxies (with higher sSFR in void galaxies), which is more prone to be affected by the LSS. Even when considering the global properties, obtained by fitting the integrated spectra, the wide field of the PMAS spectrograph let us sum over the whole extension of the galaxies, while other surveys might be too restricted to the innermost parts of galaxies.

The lack of consensus when addressing observationally the sSFR in voids and in denser environments may be caused by the different definitions of LSS, as well as SFR tracers. We plan to tackle this issue from a different point of view in a future work, calculating the SFR from emission line fluxes, to add more arguments to the debate.

\subsection{ Mass assembly, ex situ accretion of stars, and mergers in voids}

The stellar properties that we observe today depend on how the galaxies have formed and accreted mass along their life. The stellar mass in a galaxy can come from star formation that happened inside the galaxy itself, or from accretion of stars from other galaxies that have been merged. Both of these processes may depend on the LSS environment.

The radial distribution of the stellar mass surface density carries information on about how stars have formed or have migrated to their current observed radii (e.g. \citealt{Roskar2008}), and reflects the mass growth at different distances. We can also obtain information about the spatial growth of galaxies through the age profiles. Their negative gradient is a clear evidence of the inside-out formation of galaxies \citep{Perez2013}: the innermost regions form earlier, and then the rest of the disc is built at later epochs. In the case of ellipticals, their central core formed at $z\leq2$, and then they built most of their mass feeding on minor mergers \citep{Oser2010}. 

It is easier to trace the mass accretion histories through simulations than with observed data. They do so in \citet{Rodriguez-Medrano2024}, using the Illustris TNG300 simulation \citep{Nelson2019, Pillepich2018}. They find that the mass accretion has happened more recently in voids than in other environments. Mergers (ex situ processes) are one of the main drivers of mass accretion in galaxies. It has been shown that although the number of mergers does not vary between different environments, they happen at later epochs in voids with respect to denser environments \citep{Rosas-Guevara2022, Rodriguez-Medrano2024}.

In this work, we compare the stellar mass surface density profiles of the galaxies in cosmic voids to those of our control sample, and find that globally, galaxies in voids are less dense than their counterparts. This happens especially for ellipticals, although we must remember that it is also this morphological type that has the biggest shift in stellar mass. One could attribute this shift in the stellar mass density to mergers being more common in filaments and walls (specifically, dry mergers, as the morphological types that are more different in density are not in age). However, as it was already discussed, the number of mergers in both environments has been found to be equivalent.

We also find differences in the radial profiles, not only in terms of the values they attain (consistently more separated in elliptical galaxies), but also in their shapes. Specifically, it appears that the innermost regions of earlier-type galaxies (E, S0, and Sa) deviate from the typical profile by exhibiting lower densities. This observation suggests that the bulges, predominantly present in galaxies of these morphological types, are less dense in voids. This discrepancy may signify distinct formation and evolutionary processes. This finding aligns with our HLR measurements: bulges of lower stellar mass surface density (and therefore less surface brightness) mean higher values of HLR. This, along with the shift towards more important star formation in ETG in voids, is a sign of larger inner discs (or smaller bulge-to-disc ratio, B/D). This parameter is hard to determine for elliptical galaxies, and would need the use of image decomposition techniques.

\section{Summary and conclusions} \label{Section: Conclusions}

We have analysed 118 IFU datacubes of galaxies in voids from the CAVITY project, observed with the spectrograph PMAS/PPaK and the 3.5m telescope in the Calar Alto Observatory. We have obtained their stellar population properties through fitting their continuum with the full spectral fitting algorithm \starlight \ and making use of the \pycasso \ pipeline to preprocess the data and handle the 3D results.

We obtained the total stellar mass, stellar mass density, light-weighted ages, SFR, and sSFR as a result of the spectral fitting. For the spatially resolved information, we calculated the HLR, position angle, and ellipticity of each galaxy, to retrieve the radial profiles from the maps of said stellar properties. They were estimated by taking the median value of elliptical annuli growing in units of R/HLR, centred in the centre of the galaxy. We then stacked the profiles of galaxies of each morphological type, and in bins of stellar mass and morphology, to assess their overall behaviour.

We repeated the same methodology with a control sample of galaxies in the filament and walls from the CALIFA survey, removing those that belong to groups of 30 or more members following the \citet{Tempel2017} catalogue or those that stay inside voids as listed in the \citet{Pan2012} catalogue. We then took pairs of galaxies of the same morphological type and the closest total stellar mass, resulting in a control sample of 118 filament galaxies with the most similar distribution in morphological type and mass.

Our main results, based on the global and spatially resolved analysis, are as follows:
\begin{itemize}
  \item The void galaxies in our selection are slightly more extended (have a higher HLR) than those in the control sample, with an overall difference of $\Delta_\text{void-fil}$(log HLR) = $0.05\pm0.20$, mainly due to the galaxies in the mass range $10^{9.5}$ to $10^{10}$ M$_\odot$, which are more extended in voids.
  \item Galaxies in voids have a lower stellar mass surface density than galaxies in filaments and walls, with\ an overall difference of $\Delta_\text{void-fil}(\text{log }\Sigma_\star)=-0.17\pm0.24$. These differences mostly occur in elliptical galaxies, which is consistent with what appears in the radial profiles. We find that the central regions of early spiral and lenticular galaxies, probably dominated by the bulge component, have a lower stellar mass density in voids. We also find the outer parts of spirals to be less dense for galaxies in voids.
  \item Galaxies in voids are younger than those in the control sample. Their global light-weighted ages vary in general by $\Delta_\text{void-fil}(\langle \text{log t}\rangle _L)=-0.19\pm0.41$. These differences mainly appear in the discs of spiral galaxies, as well as for Sas everywhere outside of the central region (R/HLR > 0.3).
  \item The SFR is slightly higher in galaxies than in voids, by $\Delta_{\text{void-fil}}$(log SFR) = $0.15\pm0.72$, especially in early types. However, we do not find this difference to be statistically significant enough except for Sas. The amount of quenched ETG in filaments and walls is higher, indicating a slower transition from star-forming to quiescent in void galaxies.
  \item The sSFR is also slightly higher in voids for all morphological types, with an overall median deviation of $\Delta_\text{void-fil}$(log sSFR) = $0.17\pm0.72$. Similarly to the SFR, the differences in the global sSFR are only statistically significant for Sas. Through the radial profiles, we find that the values reached for galaxies in voids stay above those of galaxies in filament and walls for all galactocentric distances for ETG. We also find a mismatch in the spiral types, which have larger values of sSFR in their outer parts, probably more associated with the disc.
\end{itemize}

We find that, consistently with the results found in the literature, galaxies in voids are affected by their surroundings, which makes them evolve slower than in higher density environments, especially in their outer parts. The extent of the influence of the void environment is stronger for low-mass galaxies. We find that after a certain mass threshold is reached (between 10$^{10.5}$ and $10^{11.0}$ M$_\odot$), the internal processes become more important when driving galactic evolution. We do not find differences in the fraction of quenched galaxies, probably because of the way we chose our control sample. Differences in the evolutionary properties of void galaxies can be due to mechanisms such as more gas cold mode accretion, a higher halo mass-to-stellar mass ratio, or different paths for mass assembly and bulge formation.

Due to the selection made for the void sample, as explained in Section \ref{Section: selection voids}, we are only studying the galaxies in the brightest end of the colour-magnitude diagram. This could mean that many of the properties we are obtaining could not be representative of all galaxies in voids. However, due to the way we have constructed the control sample, the galaxies we are comparing are biased in the same way, so the selection effect should not have any repercussions on the comparison between them. We plan to carry out a more complete study once we have more data, to diminish the effect of the selection bias as well as to have more statistics to get more robust results.

\begin{acknowledgements} 
\balance
A.C., R.G.D., and R.G.B. acknowledge financial support from the Severo Ochoa grant CEX2021-001131-S funded by MCIN/AEI/ 10.13039/501100011033, and PID2019-109067GB-I00, and PID2022-141755NB-I00. 
Paper based on observations collected at Centro Astron\'omico Hispano en Andaluc\'ia (CAHA) at Calar Alto, operated jointly by Instituto de Astrof\'isica de Andaluc\'ia (CSIC) and Junta de Andaluc\'ia. 
This paper is based on data obtained by the CAVITY Survey, funded by the Spanish Ministery of Science and Innovation under grant PID2020-113689GB-I00. 
We acknowledge financial support by the research projects
AYA2017-84897-P
and PID2020-114414GB-I00, financed by MCIN/AEI/10.13039/501100011033, the project A-FQM-510-
UGR20 financed from FEDER/Junta de Andaluc\'ia-Consejer\'ia de Transforamci\'on Econ\'omica, Industria, Conocimiento y Universidades/Proyecto and by
the grants P20\_00334 and FQM108, financed by the Junta de Andaluc\'ia
(Spain).
TRL acknowledges support from Juan de la Cierva fellowship (IJC2020-043742-I), financed by MCIN/AEI/10.13039/501100011033. 
LSM acknowledges support from Juan de la Cierva fellowship (IJC2019-041527-I) and from project PID2020-114414GB-100, financed by MCIN/AEI/10.13039/501100011033. 
SDP acknowledges financial support from Juan de la Cierva Formaci\'on fellowship (FJC2021-047523-I) financed by MCIN/AEI/10.13039/501100011033 and by the European Union `NextGenerationEU'/PRTR, Ministerio de Econom\'ia y Competitividad under grants PID2019-107408GB-C44, PID2022-136598NB-C32, and is grateful to the Natural Sciences and Engineering Research Council of Canada, the Fonds de Recherche du Qu\'ebec, and the Canada Foundation for Innovation for funding.
DE acknowledges support from a Beatriz Galindo senior fellowship (BG20/00224) from the Spanish Ministry of Science and Innovation,   
and project P20-00334  financed by the Junta de Andaluc\'{i}a. 
M.A-F. acknowledges support from the Emergia program (EMERGIA20\_38888) from Consejer\'ia de Universidad, Investigaci\'on e Innovaci\'on de la Junta de Andaluc\'ia.
G.B-C acknowledges financial support from grants PID2020-114461GB-I00
and CEX2021-001131-S, funded by MCIN/AEI/10.13039/501100011033, from
Junta de Andaluc\'ia (Spain) grant P20-00880 (FEDER, EU) and from grant
PRE2018-086111 funded by MCIN/AEI/10.13039/501100011033 and by 'ESF
Investing in your future'.
GTR acknowledges financial support from the research project PRE2021-098736, funded by MCIN/AEI/10.13039/501100011033 and FSE+.
I.M.C. acknowledges support from the Spanish Ministry of Universities ref. UNI/551/2021-May 26 (EU Next Generation funds), and from ANID programme FONDECYT Postdoctorado 3230653. 
JFB acknowledges support from the PID2022-140869NB-I00 grant from the Spanish Ministry of Science and Innovation.
H.M.C acknowledge support from IUF and CNES.
AFM acknowledges support from grants RYC2021-031099-I and PID2021-123313NA-I00 of MICIN/AEI/10.13039/501100011033/FEDER,UE,NextGenerationEU/PRT.
JR acknowledges funding from University of La Laguna through the Margarita Salas Program from the Spanish Ministry of Universities ref. UNI/551/2021-May 26, and under the EU Next Generation.
We acknowledge the helpful comments made by Petra Awad to improve this work.
This research made use of {\sc astropy}, a community-developed core {\sc python} (\url{http://www.python.org}) package for Astronomy \citep{Astropy2013}; 
{\sc matplotlib} \citep{matplotlib2007}; {\sc numpy} \citep{numpy2011}; {\sc scipy} \citep{scipy2001}; {\sc pandas} \citep{pandas2010}; and {\sc seaborn} \citep{seaborn2017}.
This research has made use of the NASA/IPAC Extragalactic Database, operated by the Jet Propulsion Laboratory of the California Institute of Technology, un centract with the National Aeronautics and Space Administration. Funding for SDSS-III has been provided by the Alfred P. Sloan Foundation, the Participating Institutions, the National Science Foundation, and the U.S. Department of Energy Office of Science. The SDSS-III Web site is \url{http://www.sdss3.org/}. The SDSS-IV site is \url{http://www.sdss.org}.
\end{acknowledgements}

\bibliographystyle{aa}
\bibliography{bibliography}

\clearpage
\balance
\begin{appendix}

\onecolumn

\section{Table of global properties} 

Table \ref{Tab: tabla anexo} has the values of the global properties of each CAVITY galaxy analysed in this work. These are the stellar population properties (total stellar mass, stellar mass surface density, light-weighted ages, SFR and sSFR), the morphological type and the half-light radius. Quiescent galaxies (those with SFR = 0), have empty values of log SFR and log sSFR. This catalogue will be available with the first CAVITY DR\footnote{\label{urlcavity}\url{https://cavity.caha.es/}}.

\section{2D maps} \label{section: appendix 2D maps}

We also share the 2D maps of the stellar populations of the CAVITY galaxies within the DR in the CAVITY data base\footref{urlcavity}: the stellar mass density, light-weighted age, SFR and sSFR. The pixels with S/N lower than 3 are masked, as well as other sources in the field of view, and the maps are segmented in Voronoi zones to reach a target S/N of 20.  

\begin{figure}[h]
\centering
\includegraphics[width=\textwidth]{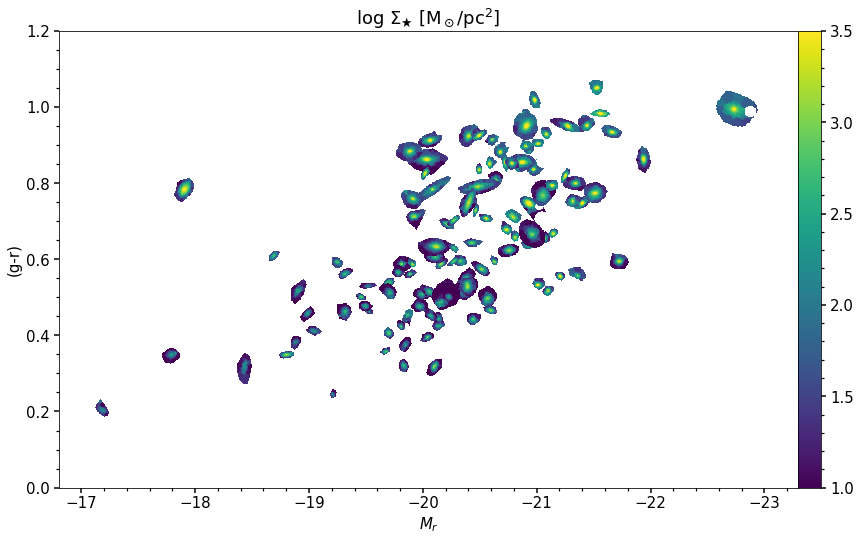}
\caption{2D maps of the stellar mass surface density of every void galaxy in our sample, placed in their respective location in the colour-magnitude diagram. The maps are oriented with the north up and the east to the left.}
\label{fig:maps_color_mag} 
\end{figure}

In Figure \ref{fig:maps_color_mag}, we show the maps of the present stellar mass surface density of every galaxy, located in the colour-magnitude diagram according to their global $M_r$ and $g - r$ values. The irregular shapes in some maps are due to the spatial masking of other sources present in the datacube, done previous to the fitting (and the creation of the Voronoi zones). It is clearly seen that galaxies with larger values of log $\Sigma_\star$ populate the brighter and redder part of the diagram. Comparing to the colour-magnitude diagram coloured by morphological type in Figure \ref{fig:col-mag}, we can check that these galaxies correspond to earlier types. 

\begin{figure}
\centering
\includegraphics[width=0.45\textwidth]{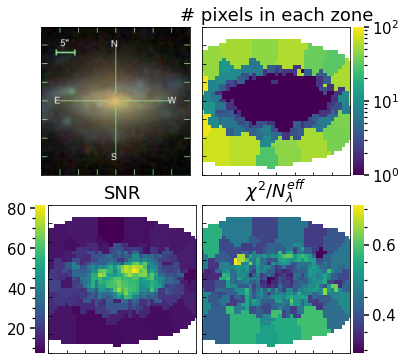}
\caption{Maps with fit quality indicators of CAVITY54706.} 
\label{fig:QC_maps}
\end{figure}

Figure \ref{fig:QC_maps} shows the quality control maps of one void galaxy, CAVITY54706. Top left plot is an image in the r-band of the galaxy obtained from SDSS. On its right, there is the map with the Voronoi zones. The colouring marks the number of pixels that belong to each zone, that will be summed together to reach the S/N threshold value. Bottom left panel shows the S/N of each zone. We can see that the spaxels are aggregated on the outer parts of the galaxy, where there is a lower S/N. The last plot has the information about the quality of the fit in each zone, through the $\chi^2$ divided by the number of effective spectral pixels (the total minus the clipped during the fitting process by the algorithm).

\begin{figure*}
\centering
\includegraphics[width=\textwidth]{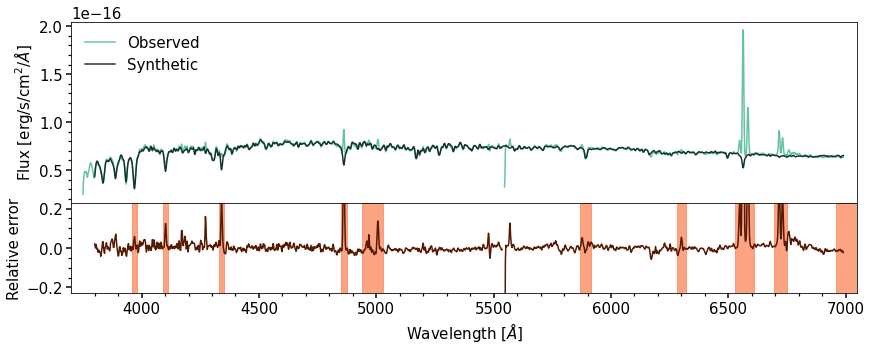}
\caption{Example of a \starlight \ spectral fit. \textit{Top panel}: Observed and synthetic spectra of the central pixel of CAVITY54706. \textit{Bottom panel}: Relative error of the fit: the observed flux minus the synthetic, divided by the observed flux in the normalisation wavelength ($5635\AA$). Wavelength sections in orange are masked during the fit.} 
\label{fig:fit_spectrum}
\end{figure*}

An example of a fit can be seen in Figure \ref{fig:fit_spectrum}. The top panel has the observed (in green) and synthetic (in darker green) spectra of one spaxel of a galaxy. The bottom panel shows the relative residual spectrum, calculated as the division between the difference $f_{\text{obs}}-f_{\text{syn}}$ and the observed flux in the normalisation wavelength, $f_{\text{obs,norm}}$. The orange sections mark in which wavelength windows the spectrum was masked during the fit.

\begin{longtable}[p]{c|ccccccc}
\centering
 & Hubble &  log M$_\star$ &  log $\Sigma_\star$ &  $\langle \text{log t}\rangle _L$ &  log SFR &  log sSFR &   HLR \\
 & Type &  [M$_\odot]$ &  [M$_\odot/$pc$^2]$ &  [yr] &  [M$_\odot$/yr] &  [Gyr$^{-1}]$ &   kpc \\  \hline
\endfirsthead

 & Hubble &  log M$_\star$ &  log $\Sigma_\star$ &  $\langle \text{log t}\rangle _L$ &  log SFR &  log sSFR &   HLR \\
 & Type &  [M$_\odot]$ &  [M$_\odot/$pc$^2]$ &  [yr] &  [M$_\odot$/yr] &  [Gyr$^{-1}]$ &   kpc \\  \hline
\endhead
\multicolumn{8}{r}{{Continued on next page}} \\
\endfoot

\endlastfoot
CAVITY10595 &    S0 &    11.08 &      2.71 &     9.95 &   -0.38 &    -2.46 &  3.26 \\
CAVITY11248 &    Sc &    10.30 &      2.08 &     8.53 &    0.71 &    -0.58 &  4.04 \\
CAVITY12190 &    Sc &    10.35 &      2.02 &     8.81 &    0.73 &    -0.62 &  4.95 \\
CAVITY16768 &    Sc &    10.31 &      2.03 &     8.35 &    0.73 &    -0.58 &  3.95 \\
CAVITY16881 &    Sa &    10.82 &      2.52 &     9.24 &    0.53 &    -1.29 &  3.09 \\
CAVITY17302 &    Sb &    10.40 &      2.12 &     8.87 &   -0.03 &    -1.43 &  4.49 \\
CAVITY17344 &    Sc &    10.77 &      2.43 &     9.34 &    0.29 &    -1.48 &  3.06 \\
CAVITY17370 &    Sc &     9.37 &      1.52 &     7.94 &    0.14 &    -0.23 &  3.27 \\
CAVITY17616 &    S0 &    11.23 &      2.57 &     9.82 &     -- &      -- &  3.60 \\
CAVITY18857 &    Sb &    10.32 &      2.00 &     9.05 &    0.39 &    -0.94 &  5.53 \\
CAVITY18904 &    Sc &     9.94 &      1.40 &     8.34 &    0.64 &    -0.30 &  6.37 \\
CAVITY19279 &    Sb &    10.09 &      2.07 &     9.22 &   -0.03 &    -1.12 &  3.02 \\
CAVITY19959 &    Sb &    10.69 &      2.34 &     8.60 &    1.07 &    -0.61 &  4.13 \\
CAVITY20424 &    Sc &    10.50 &      2.25 &     9.10 &    0.63 &    -0.87 &  3.48 \\
CAVITY20787 &     E &    10.80 &      2.31 &     9.68 &     -- &      -- &  3.39 \\
CAVITY27289 &    Sc &    10.08 &      1.94 &     8.74 &    0.39 &    -0.69 &  3.84 \\
CAVITY27381 &    Sc &    10.10 &      1.91 &     8.91 &    0.28 &    -0.82 &  4.20 \\
CAVITY29349 &    Sa &    10.84 &      2.50 &     9.39 &    0.44 &    -1.40 &  3.60 \\
CAVITY29867 &    Sc &    10.65 &      2.42 &     9.29 &   -1.64 &    -3.29 &  4.00 \\
CAVITY31671 &    Sb &    10.73 &      2.34 &     9.26 &    0.54 &    -1.19 &  3.46 \\
CAVITY32250 &    Sc &     9.74 &      1.49 &     8.65 &   -0.52 &    -1.25 &  7.38 \\
CAVITY32597 &    Sc &    10.11 &      2.06 &     9.12 &    0.37 &    -0.74 &  2.27 \\
CAVITY32895 &    Sc &    10.38 &      1.92 &     9.07 &    0.76 &    -0.62 &  3.22 \\
CAVITY32898 &    S0 &    10.98 &      2.54 &     9.11 &    1.07 &    -0.91 &  3.47 \\
CAVITY33678 &    Sb &     9.02 &      1.63 &     8.54 &   -0.74 &    -0.76 &  1.47 \\
CAVITY33695 &    Sc &     9.37 &      1.50 &     8.38 &    0.24 &    -0.13 &  3.73 \\
CAVITY34100 &    Sa &    10.64 &      2.44 &     9.74 &     -- &      -- &  3.53 \\
CAVITY34170 &    Sc &    10.17 &      1.73 &     8.45 &    0.33 &    -0.84 &  4.80 \\
 CAVITY3430 &     E &    10.95 &      2.52 &     9.95 &   -0.05 &    -2.00 &  3.24 \\
CAVITY35487 &    Sc &     9.85 &      1.48 &     8.02 &    0.42 &    -0.43 &  5.57 \\
CAVITY35580 &    Sc &    10.19 &      1.89 &     8.56 &    0.94 &    -0.25 &  3.37 \\
CAVITY36541 &    Sc &    10.12 &      1.89 &     8.75 &    0.29 &    -0.83 &  3.76 \\
 CAVITY3666 &    Sc &     9.72 &      1.48 &     8.41 &    0.34 &    -0.39 &  4.73 \\
 CAVITY3670 &    Sa &    10.47 &      2.32 &     8.98 &    0.75 &    -0.71 &  5.17 \\
CAVITY37527 &    Sc &     9.98 &      1.55 &     8.59 &   -0.96 &    -1.94 &  6.34 \\
CAVITY37605 &    Sc &    10.06 &      1.70 &     8.82 &    0.16 &    -0.89 &  4.33 \\
CAVITY37926 &    S0 &    10.43 &      2.14 &     9.68 &   -2.11 &    -3.54 &  1.95 \\
CAVITY37963 &    Sc &    10.18 &      1.89 &     8.53 &    0.48 &    -0.69 &  3.97 \\
CAVITY37970 &     E &    10.82 &      2.34 &     9.55 &     -- &      -- &  2.91 \\
CAVITY38659 &    Sa &    10.61 &      2.55 &     9.47 &    0.53 &    -1.08 &  2.96 \\
CAVITY40288 &     E &    10.71 &      2.25 &     9.76 &     -- &      -- &  4.29 \\
CAVITY40393 &     E &    10.87 &      2.40 &     9.76 &    0.08 &    -1.79 &  3.78 \\
CAVITY40821 &    S0 &    10.51 &      2.13 &     9.54 &     -- &      -- &  3.66 \\
CAVITY40822 &    Sc &    10.25 &      2.03 &     9.26 &   -0.58 &    -1.83 &  2.95 \\
CAVITY40825 &    Sb &    11.19 &      2.34 &     9.49 &    0.72 &    -1.47 &  7.82 \\
CAVITY41234 &    Sc &    10.51 &      2.08 &     9.26 &    0.43 &    -1.08 &  6.35 \\
CAVITY41235 &     E &    10.84 &      2.45 &     9.63 &     -- &      -- &  2.57 \\
CAVITY41359 &    Sc &    10.38 &      2.02 &     9.11 &    0.49 &    -0.89 &  6.78 \\
CAVITY41398 &    Sa &    10.81 &      2.09 &     8.99 &    0.47 &    -1.35 &  4.51 \\
CAVITY41400 &    Sa &    10.70 &      2.51 &     9.82 &     -- &      -- &  2.26 \\
CAVITY41401 &    S0 &    10.83 &      2.53 &     9.45 &    0.43 &    -1.40 &  2.31 \\
CAVITY41448 &    S0 &    11.03 &      2.40 &     9.62 &     -- &      -- &  4.73 \\
CAVITY41455 &    Sc &    10.09 &      1.68 &     8.49 &    0.18 &    -0.92 &  5.58 \\
CAVITY41495 &     E &    10.88 &      2.07 &     9.45 &     -- &      -- &  3.79 \\
CAVITY46819 &    Sb &    10.69 &      1.96 &     9.12 &     -- &      -- &  6.80 \\
CAVITY47120 &    Sc &    10.10 &      1.81 &     8.65 &    0.48 &    -0.62 &  5.32 \\
CAVITY48125 &    Sb &    10.89 &      2.47 &     9.38 &    0.44 &    -1.46 &  4.73 \\
CAVITY48361 &    S0 &    10.64 &      2.44 &     9.68 &    0.43 &    -1.21 &  3.08 \\
CAVITY49935 &    Sc &     9.45 &      1.59 &     8.81 &   -0.56 &    -1.01 &  3.34 \\
CAVITY50031 &    Sb &    10.09 &      1.72 &     8.57 &   -0.03 &    -1.12 &  3.24 \\
CAVITY50117 &    Sc &     9.73 &      1.72 &     8.64 &    0.27 &    -0.46 &  3.71 \\
CAVITY50532 &    Sc &     9.73 &      1.19 &     8.70 &   -0.76 &    -1.49 &  6.46 \\
CAVITY50943 &     E &    10.69 &      2.33 &     9.41 &   -0.03 &    -1.72 &  3.17 \\
CAVITY51102 &    Sc &    10.47 &      2.08 &     9.24 &   -0.53 &    -2.00 &  4.46 \\
CAVITY51442 &    S0 &    10.52 &      2.41 &     9.12 &    0.64 &    -0.87 &  3.33 \\
CAVITY51443 &    S0 &    10.91 &      2.53 &     9.60 &    0.32 &    -1.59 &  2.84 \\
CAVITY51933 &    Sb &    10.96 &      2.52 &     9.20 &    0.86 &    -1.09 &  4.20 \\
CAVITY52730 &    Sb &    10.95 &      2.36 &     9.57 &   -0.06 &    -2.02 &  4.36 \\
CAVITY52732 &    Sb &    10.29 &      2.01 &     9.17 &    0.46 &    -0.83 &  5.47 \\
CAVITY53259 &    S0 &    10.47 &      2.35 &     9.61 &   -0.87 &    -2.34 &  2.00 \\
CAVITY54459 &    Sb &    10.98 &      2.54 &     9.28 &    0.69 &    -1.29 &  4.05 \\
CAVITY54598 &    Sb &    11.34 &      2.87 &     9.66 &    0.91 &    -1.43 &  5.87 \\
CAVITY54706 &    Sc &    10.46 &      2.07 &     9.09 &    0.58 &    -0.89 &  3.66 \\
CAVITY55131 &    Sc &    10.60 &      1.94 &     8.88 &    0.73 &    -0.86 &  8.92 \\
CAVITY55180 &     E &    10.42 &      2.16 &     9.64 &     -- &      -- &  2.61 \\
CAVITY56289 &    Sd &     9.41 &      1.46 &     8.23 &    0.16 &    -0.24 &  3.37 \\
CAVITY56627 &    Sc &    10.02 &      1.84 &     8.63 &    0.52 &    -0.49 &  3.98 \\
CAVITY57008 &    Sa &     9.81 &      1.97 &     9.27 &   -0.14 &    -0.95 &  3.16 \\
CAVITY57232 &    Sa &    10.83 &      2.54 &     9.24 &    0.59 &    -1.24 &  3.67 \\
CAVITY58740 &    Sc &    10.17 &      1.84 &     8.58 &    0.45 &    -0.72 &  3.29 \\
CAVITY59465 &    Sb &    10.75 &      2.24 &     9.15 &    0.55 &    -1.19 &  5.24 \\
CAVITY59764 &    Sc &    10.59 &      2.04 &     8.87 &    0.30 &    -1.29 &  5.93 \\
CAVITY59902 &    Sc &    10.08 &      1.69 &     8.62 &    0.68 &    -0.41 &  4.90 \\
CAVITY59983 &    Sb &    10.43 &      2.09 &     9.13 &   -1.71 &    -3.14 &  5.73 \\
CAVITY60044 &    Sc &     9.66 &      1.69 &     8.61 &   -0.04 &    -0.70 &  3.46 \\
CAVITY60187 &    Sc &     9.87 &      1.87 &     9.16 &   -0.03 &    -0.90 &  3.37 \\
CAVITY60224 &    Sc &    10.06 &      1.67 &     8.83 &    0.07 &    -0.99 &  3.99 \\
CAVITY60322 &    Sc &    10.10 &      2.14 &     9.18 &    0.13 &    -0.97 &  2.69 \\
CAVITY60693 &    S0 &    11.06 &      2.55 &     9.97 &     -- &      -- &  3.61 \\
CAVITY61128 &     E &    10.93 &      2.35 &     9.69 &     -- &      -- &  3.62 \\
CAVITY62108 &    Sc &    10.02 &      1.73 &     8.63 &    0.21 &    -0.82 &  4.66 \\
CAVITY62262 &    Sc &     9.78 &      1.75 &     8.67 &     -- &      -- &  3.33 \\
CAVITY62323 &    Sc &    10.46 &      2.13 &     8.94 &    0.66 &    -0.80 &  4.38 \\
CAVITY62480 &    Sc &    11.01 &      2.33 &     9.56 &    0.69 &    -1.32 &  4.21 \\
CAVITY63083 &    Sc &     9.93 &      1.72 &     8.71 &     -- &      -- &  3.54 \\
CAVITY63263 &    Sc &     9.97 &      1.87 &     8.64 &    0.21 &    -0.76 &  4.59 \\
CAVITY65288 &    Sb &    10.41 &      2.08 &     9.22 &    0.67 &    -0.74 &  4.83 \\
CAVITY65303 &     E &    10.92 &      2.46 &     9.97 &    0.25 &    -1.67 &  4.88 \\
CAVITY65716 &    Sc &    10.36 &      2.15 &     8.73 &   -0.05 &    -1.41 &  4.34 \\
CAVITY65887 &    Sb &    10.49 &      2.27 &     9.42 &    0.49 &    -1.00 &  4.17 \\
CAVITY66400 &    Sb &    10.24 &      2.09 &     8.81 &    0.57 &    -0.66 &  3.10 \\
CAVITY66803 &    Sc &    10.35 &      1.77 &     8.66 &     -- &      -- &  6.21 \\
 CAVITY7906 &    Sc &     9.93 &      1.49 &     8.51 &   -0.05 &    -0.99 &  4.36 \\
 CAVITY7926 &    Sb &    10.92 &      2.62 &     8.79 &    1.16 &    -0.75 &  5.58 \\
 CAVITY8556 &     E &    10.40 &      2.18 &     9.26 &   -0.10 &    -1.49 &  4.70 \\
 CAVITY8595 &    Sb &    10.25 &      2.32 &     8.66 &    0.82 &    -0.43 &  3.55 \\
 CAVITY8645 &    Sa &    10.89 &      2.57 &     8.84 &    0.95 &    -0.94 &  3.13 \\
 CAVITY8646 &    Sa &    10.37 &      2.44 &     8.91 &    0.65 &    -0.71 &  4.19 \\
 CAVITY8766 &     E &    11.62 &      2.26 &     9.99 &     -- &      -- & 10.60 \\
VGS05 &    S0 &    10.54 &      2.15 &     9.80 &     -- &      -- &  2.57 \\
VGS31 &    Sb &    10.13 &      1.98 &     8.64 &    0.42 &    -0.71 &  2.34 \\
VGS42 &    Sc &     9.63 &      2.00 &     8.93 &   -0.40 &    -1.03 &  2.10 \\
VGS47 &    Sa &    10.69 &      2.31 &     9.38 &    0.45 &    -1.24 &  4.85 \\
VGS49 &    Sa &    10.07 &      2.10 &     8.78 &    0.54 &    -0.53 &  2.37 \\
VGS55 &    Sc &     9.57 &      1.53 &     8.48 &   -0.01 &    -0.59 &  3.33 \\
VGS56 &    Sb &    10.00 &      2.33 &     9.65 &   -0.45 &    -1.45 &  2.38 \\
VGS57 &    S0 &    10.52 &      2.18 &     9.18 &    0.34 &    -1.18 &  3.10 \\
VGS58 &    Sc &     9.02 &      1.50 &     8.10 &   -0.30 &    -0.32 &  1.51 \\
\caption{Global stellar population and structural properties of each CAVITY galaxy in our sample.} \label{Tab: tabla anexo} 
\end{longtable}

\end{appendix}
\end{document}